\documentclass[fleqn,usenatbib]{mnras}

\usepackage[T1]{fontenc}
\usepackage{ae,aecompl}

\usepackage{graphicx}	
\usepackage{amsmath}	
\usepackage{amssymb}	

\newcommand{\MP}{\textit{M}$_{\textup{p}}$} 
\newcommand{\RP}{\textit{R}$_{\textup{p}}$} 
\newcommand{\MJ}{\textit{M}$_{\textup{J}}$} 
\newcommand{\RJ}{\textit{R}$_{\textup{J}}$} 

\newcommand{\appx}{\arcsec\,pixel$^{-1}$\,} 

\newcommand{\bjdtdb}{\ensuremath{\rm {BJD_{TDB}}}}
\newcommand{\teff}{\textit{T}$_{\textup{eff}}$}
\newcommand{\feh}{\ensuremath{\left[{\rm Fe}/{\rm H}\right]}}

\newcommand{\fave}{\langle F \rangle}
\newcommand{\fluxcgs}{10$^9$ erg s$^{-1}$ cm$^{-2}$}
\newcommand{\ecosw}{\ensuremath{e\cos{\omega_*}}}
\newcommand{\esinw}{\ensuremath{e\sin{\omega_*}}}

\newcommand{\kms}{\,km\,s$^{-1}$} 
\newcommand{\ms}{\,m\,s$^{-1}$} 
\newcommand{\msd}{\,m\,s$^{-1}$\,d$^{-1}$} 

\title[KELT-10b: A Hot Sub-Jupiter Transiting a V$=$10.7 Early G-Star]{KELT-10b: The First Transiting Exoplanet from the KELT-South Survey -- A Hot Sub-Jupiter Transiting a~V$=$10.7 Early G-Star}

\author[Rudolf B.~Kuhn et al.]
{
Rudolf~B.~Kuhn,$^{1}$\thanks{E-mail: rudi@saao.ac.za}
Joseph~E.~Rodriguez,$^{2}$
Karen~A.~Collins,$^{2,3}$
Michael~B.~Lund,$^{2}$
\newauthor
Robert~J.~Siverd,$^{4}$
Knicole~D.~Col\'on,$^{5,6,7}$
Joshua~Pepper,$^{5}$
Keivan~G.~Stassun,$^{2,8}$
\newauthor
Phillip~A.~Cargile,$^{9}$
David~J.~James,$^{10}$
Kaloyan~Penev,$^{11}$
George~Zhou,$^{12}$
\newauthor
Daniel~Bayliss,$^{13,14}$
T.G.~Tan,$^{15}$
Ivan~A.~Curtis,$^{16}$
Stephane~Udry,$^{13}$
Damien~Segransan,$^{13}$
\newauthor
Dimitri~Mawet,$^{17,18}$
Saurav Dhital,$^{19}$
Jack~Soutter,$^{20}$
Rhodes~Hart,$^{20}$
Brad~Carter,$^{20}$
\newauthor
B.~Scott~Gaudi,$^{21}$
Gordon~Myers,$^{22,23}$
Thomas~G.~Beatty,$^{24,25}$
Jason~D.~Eastman,$^{9}$
\newauthor
Daniel~E.~Reichart,$^{26}$
Joshua~B.~Haislip,$^{26}$
John~Kielkopf,$^{3}$
Allyson~Bieryla,$^{9}$
\newauthor
David~W.~Latham,$^{9}$
Eric~L.~N.~Jensen,$^{27}$
Thomas~E.~Oberst,$^{28}$
and Daniel~J.~Stevens$^{21}$
\vspace{0.3cm}\\
Author affiliations are listed on the final page.
}
\date{Accepted XXX. Received YYY; in original form ZZZ}

\pubyear{2015}

\begin{document}
\label{firstpage}
\pagerange{\pageref{firstpage}--\pageref{lastpage}}
\maketitle

\begin{abstract}
We report the discovery of KELT-10b, the first transiting exoplanet discovered using the KELT-South telescope. KELT-10b is a highly inflated sub-Jupiter mass planet transiting a relatively bright $V = 10.7$ star (TYC 8378-64-1), with \teff~=~$5948\pm74$~K, $\log{g}$~=~$4.319_{-0.030}^{+0.020}$ and \feh~=~$0.09_{-0.10}^{+0.11}$, an inferred mass $M_{*}$~=~$1.112_{-0.061}^{+0.055}$~$M_{\sun}$ and radius $R_{*}$~=~$1.209_{-0.035}^{+0.047}$~$R_{\sun}$. The planet has a radius \RP~=~$1.399_{-0.049}^{+0.069}$~\RJ\space and mass \MP~=~$0.679_{-0.038}^{+0.039}$~\MJ. The planet has an eccentricity consistent with zero and a semi-major axis $a$~=~$0.05250_{-0.00097}^{+0.00086}$~AU. The best fitting linear ephemeris is $T_{0}$~=~2457066.72045$\pm$0.00027\space\bjdtdb\space and P~=~4.1662739$\pm$0.0000063 days. This planet joins a group of highly inflated transiting exoplanets with a radius larger and a mass less than that of Jupiter. The planet, which boasts deep transits of 1.4\%, has a relatively high equilibrium temperature of \textit{T}$_{\textup{eq}}$~=~$1377_{-23}^{+28}$~K, assuming zero albedo and perfect heat redistribution. KELT-10b receives an estimated insolation of $0.817_{-0.054}^{+0.068}$~$\times$~\fluxcgs, which places it far above the insolation threshold above which hot Jupiters exhibit increasing amounts of radius inflation. Evolutionary analysis of the host star suggests that KELT-10b may not survive beyond the current subgiant phase, depending on the rate of in-spiral of the planet over the next few Gyr. The planet transits a relatively bright star and exhibits the third largest transit depth of all transiting exoplanets with V $<$ 11 in the southern hemisphere, making it a promising candidate for future atmospheric characterization studies.
\end{abstract}


\section{Introduction}
\label{sec:Introduction}

Ground-based searches for transiting exoplanets have to date found more than 150 exoplanets \citep{Schneider:2011}.\footnote{http://exoplanet.eu/} Small aperture, wide field robotic telescopes used by survey groups like HATNet and HATSouth \citep{Bakos:2004, Bakos:2013}, SuperWASP and WASP-South \citep{Pollacco:2006, Hellier:2011}, XO \citep{McCullough:2005} and TrES \citep{Alonso:2004} have been the most successful at finding exoplanets. Even with the success of these surveys, fewer than 25 transiting exoplanets with V $<$ 11 have been discovered from the southern hemisphere. Planets transiting bright stars are potentially the most scientifically valuable, as these systems provide the opportunity to determine many precise properties of the planet itself (see \citeauthor{Winn:2009} \citeyear{Winn:2009}).

To illustrate this point, consider the transiting planet HD209458b \citep{Charbonneau:2000,Henry:2000}. With a magnitude of V~=~7.95, this is one of the brightest transiting planet host stars and thus one of the best studied exoplanets. To date, over 50 papers on various properties of the exoplanet\footnote{See http://exoplanet.eu/catalog/hd\_209458\_b/ for a complete list of all publications.} have been published. Currently HD219134b \citep{Motalebi:2015} is the nearest and brightest host star with a known transiting exoplanet. This system shows enormous promise for future follow-up observations, which would better constrain the system architecture and characterize the physical properties of the planet. 

More information can be gathered from  bright transiting exoplanets due to the greater amount of flux from the host stars, enabling high-precision follow-up and characterization. It is for this specific reason the \textbf{T}ransiting \textbf{E}xoplanet \textbf{S}urvey \textbf{S}atellite (TESS) \citep{Ricker:2015} will perform a wide-field survey for planets that transit bright host stars. The TESS team expects to find $\sim$1700 transiting exoplanets in the lifetime of the mission \citep{Sullivan:2015} and a large fraction of TESS discovered planets will be attractive targets for follow-up measurements and atmospheric characterization due to their brightness.

The \textbf{K}ilodegree \textbf{E}xtremely \textbf{L}ittle \textbf{T}elescope-South (KELT-South) transit survey \citep{Pepper:2012} was built to detect giant, short-period transiting planets orbiting bright stars. KELT-South was constructed to detect transiting planets orbiting stars with 8 $<$ V $<$ 10 \citep{Pepper:2003}, since most RV surveys target stars brighter than V~=~9 and most transit surveys saturate for stars brighter than V~=~10. Although the original specification was for V $<$ 10, the KELT telescopes are capable of finding exoplanets around fainter stars and the first of the KELT telescopes, KELT-North \citep{Pepper:2007}, has already had success in finding brown dwarfs and planets around a large magnitude range of stars. KELT-1b \citep{Siverd:2012} is a 27 \MJ\space brown dwarf transiting a $V$ = 10.7 F-star, with KELT-1 being the brightest star known to host a transiting brown dwarf. KELT-2Ab \citep{Beatty:2012} is a hot Jupiter transiting the bright $V$ = 8.77 primary star in a visual binary system. KELT-3b \citep{Pepper:2013} is a hot Jupiter transiting a $V$ = 9.8 slightly evolved late F-star. KELT-4Ab \citep{Eastman:2016} is an inflated hot Jupiter transiting the brightest component in a $V$ = 9.98 hierarchical triple. KELT-6b \citep{Collins:2014} is a mildly-inflated Saturn-mass planet transiting a metal poor, slightly evolved $V$ = 10.38 late F-star. KELT-7b \citep{Bieryla:2015} is a hot Jupiter transiting a bright $V$ = 8.54 rapidly rotating F-star. KELT-8b \citep{Fulton:2015} is a highly inflated hot Jupiter, with \RP = 1.86$_{-0.16}^{+0.18}$ \RJ, transiting a relatively bright $V$ = 10.83 G-star. This makes KELT-8b one of the best candidates for atmospheric characterization with transmission spectroscopy. 

In this paper we describe the discovery and properties of KELT-10b, a hot sub-Jupiter (planets with radii similar to Jupiter and mass less than Jupiter) transiting the relatively bright $V$~=~10.7 early G-star TYC 8378-64-1, the first exoplanet discovered by the KELT-South survey.

\section{The KELT-South survey}
\label{sec:KS_Survey}
Because this is the first paper describing an exoplanet discovery from the KELT-South survey, we describe the telescope instrumentation and observing site, survey strategy, data reduction methodology, data reduction pipeline, light curve combination and candidate exoplanet selection criteria in detail. Additional information is available in \S2 of \citet{Siverd:2012}.

\subsection{KELT-South instrumentation and observing site}
\label{sec:KS_Telescope}
Although the KELT-South instrumentation paper \citep{Pepper:2012} describes the KELT-South hardware in detail, there have been a number of small improvements and reconfigurations since that publication. In this section we summarise the current KELT-South hardware and computational setup and provide some basic performance metrics.

The KELT-South telescope consists of 3 main parts; camera and lens, mount, and control computer. The KELT-South detector is an Apogee Instruments Alta U16M thermo-electrically cooled CCD camera. The camera uses the Kodak KAF-16803 front illuminated CCD with 4096 $\times$ 4096 9 $\mu$m pixels (36.88 $\times$ 36.88 mm detector area) and has a peak quantum efficiency of $\sim$ 60\% at 550 nm. The device is read out at 16 bit resolution at 1 MHz, giving a full-frame readout time of $\sim$ 30 s. The CCD camera is maintained at -20$^\circ$ C to reduce random thermal noise. The typical system noise for the CCD is given by the manufacturer as $\sim$ 9 $e^{-}$ at 1 MHz readout speed. In laboratory testing the nominal dark current was $<$ 1.4 e$^{-}$ pixel$^{-1}$ s$^{-1}$ at a temperature of -20$^\circ$ C and the CCD has temperature stability of $\pm$0.1$\degr$ C. The CCD operates at a conversion gain of 1.4 $e^{-}$ ADU$^{-1}$ and the full-well depth is listed as $\sim$ 100000 e$^{-}$, but the analogue-to-digital converter (ADC) saturates at 65535 ADU ($\sim$ 92000 $e^{-}$). The linear dynamic range is listed as 80 dB and the photoresponse non-linearity and non-uniformity are given as 1\%. The CCD also has anti-blooming protection to prevent image bleed from over-exposed regions.

KELT-South uses a Mamiya 645 80 mm f/1.9 medium-format manual focus lens with a 42 mm aperture. The field of view using this lens is 26 deg $\times$ 26 deg and provides $\sim$~23\appx image scale. In front of the lens is a Kodak Wratten No. 8 red-pass filter. This filter has a 50\% transmission point at $\sim$ 490 nm. All the optical components are mounted on a Paramount ME Robotic Telescope mount manufactured by Software Bisque. The Paramount ME is a research-grade German Equatorial Mount design and has a tracking error of $\pm$ 5$\arcsec$. The telescope needs to perform a meridian flip when taking images for fields crossing the meridian due to the way in which the camera and lens are attached to the mount.

The Dell Optiplex 755 small-form-factor computer that controls all aspects of the telescope operation runs the Windows XP operating system and is housed in a temperature controlled cabinet manufactured by Rittal. The Rittal cabinet also houses the UPS (uninterruptible power supply) unit to protect the computer against voltage spikes and brief power outages. The software packages (TheSky 6 Professional, CCDSoft and TPoint) provided by Software Bisque enable the control computer to operate the CCD and mount via a script-accessible interface. Various scripts, written in Visual Basic Scripting Edition, are used to perform the observing procedures, basic image analysis to eliminate bad images and data archiving.

The telescope is located at the South African Astronomical Observatory site (20\degr38\arcmin48\arcsec\space E, 32\degr22\arcmin46\arcsec\space S, Altitude 1768 m) near Sutherland, South Africa.  Wind speeds less than 45 \kms\space occur 90\% of the time throughout the year and the median relative humidity level is 45\% (includes day and night time). 

The point-spread function (PSF) of a star is dependent on the position of the star on the KELT-South CCD sensor and is characterized by a full width at half maximum (FWHM) value of between 3 and 6 pixels. The KELT-South telescope operates slightly defocused to avoid under-sampling of the stars. The telescope focus is not adjusted on a nightly basis and is kept at a fixed position throughout the lifetime of the telescope. Strict photometric conditions are required for optimal photometry, but a benefit of the large pixel scale of KELT-South is that large atmospheric seeing variations do not affect our ability to observe and good seeing conditions are not necessary for 10 millimag precision photometry for the brightest stars in the KELT-South survey.

\subsection{Survey strategy}
\label{sec:KS_SurveyStragety}
In the regular survey mode the telescope observes a number of fields located around the sky throughout the night. At present there are 29 fields observed in the survey schedule, which cover $\sim$ 50\% of the southern sky. Survey observations consist of 150 s exposures with a typical per-field cadence of 15 to 30 min. KELT-South has been collecting survey data since 2010 February and to date has acquired between 9000 and 16000 images per field. Given this quantity of data and the typical achieved photometric precision of $\sim$ 1\% for all stars with \textit{V} $\leq$ 11, the KELT-South survey is able to detect short-period giant transiting exoplanets orbiting most FGK dwarf stars with magnitudes from saturation near \textit{V} $\sim$ 8 down to \textit{V} $\sim$ 11.

\subsection{Data reduction}
\label{sec:KS_Data_Reduction_Pipeline}
When constructing a data reduction pipeline, there are generally three options; aperture photometry, PSF fitting photometry and image subtraction. Although aperture photometry and PSF photometry have been in use for an extended period of time and both of these reduction techniques are very well tested, KELT-South uses an image subtraction technique, first proposed by \cite{Tomaney:1996}. Image subtraction has been shown to work much better than aperture/PSF photometry at finding variable stars and transiting exoplanets in extremely crowded fields like globular clusters \citep{Olech:1999, Hartman:2008, McCormac:2014} and open clusters \citep{Hartman:2004, Howell:2005, Montalto:2007}. The KELT-South telescope has an extremely wide field of view and large pixel scale, which necessitated the choice of an image subtraction data reduction pipeline over the other options.

KELT-South shares a data reduction pipeline with the KELT-North telescope \citep{Pepper:2003,Gaudi:2015}. Both telescopes use similar optics, with the only difference between the telescopes being the detector and minor differences in the observing procedures. It is thus easy to use the same pipeline and reduction procedures for both telescopes with minor changes to accommodate these differences. KELT-South uses a slightly updated version of the pipeline, with new routines that are able to identify more individual stars in extremely crowded regions. Additional information is available in \citet{Siverd:2012} and \citet{Pepper:2012}. 

\subsection{Pipeline overview}
\label{sec:KS_Pipeline}
The KELT project makes use of a heavily modified \textsc{ISIS}\footnote{See http://www2.iap.fr/users/alard/package.html} difference-image-analysis package \citep{Alard:1998, Alard:2000, Hartman:2004} to achieve high-precision relative photometry. Raw data images are dark-subtracted and flat-fielded. A master dark image is acquired by median combination of hundreds of dark images taken at the start of the observing season. This dark frame is used for all data reduction and is infrequently updated. The data reduction pipeline is able to use a master dark image in this manner as we have found that the Poisson noise due to the bright background sky is much larger than the systematic errors introduced by using an outdated dark image. A master flat field image is used for all science images, which was constructed using hundreds of twilight sky flats, each of which was individually bias-subtracted, scaled-dark-subtracted, and additionally gradient-corrected prior to combination. 

Light curves for individual objects are then constructed using the heavily modified \textsc{ISIS} image-subtraction pipeline. Image subtraction is highly computer intensive. To improve performance the \textsc{ISIS} scripts were modified to facilitate distributed image reduction across many computers in parallel. Other elements of the \textsc{ISIS} package were also modified or replaced with faster alternatives. For example, the standard \textsc{ISIS} source-identification routines are ill equipped to deal with the nature and ubiquity of the aberrations in KELT-South images and the \textsc{extract} utility was replaced with the \textsc{sextractor} program \citep{Bertin:1996}. More details of these modifications can be found online.\footnote{http://astro.phy.vanderbilt.edu/$\sim$siverdrj/soft/is3/index.html} 

Each observational image is examined for pointing scatter and a suitable high-quality image is selected to serve as an astrometric reference for that field. All the other images of that field are then registered (aligned) to this image. Shifts in \textit{x} and \textit{y} positions of the individual images are caused by incorrect pointing of the telescope or slight drifts due to an incorrect pointing model. Once all the images are registered, we median-combine all images, and create a list of all images that had a large number of pixels with outlying values from the median. We then median-combine again all images not on this list, and repeat the process of culling images two more times until the final median image is based on just the high-quality images. The highest quality images used to construct the master reference image are typically (a) acquired at low air mass, and (b) have low sky background flux. The result of this process is a maximally high signal-to-noise ratio (SNR) image that we use to define positions and fluxes of all objects identified for extraction.

The reference image is convolved using a series of Gaussian functions to match the object shapes and fluxes for each individual image, and the convolved reference is then subtracted from that image. By first matching object shapes in this fashion, we ensure that any residual flux has the same shape (PSF) as the original image. The residual flux from the subtracted image is then measured using PSF-weighted aperture photometry and added to the median flux of the object identified on the reference image. Median fluxes are obtained from the reference image using the stand-alone \textsc{DAOPHOT II}  \citep{Stetson:1987, Stetson:1990}. This allows us to assemble light curves for each individual object. Light curves are generated over the total baseline of the observations, rather than in separate time segments such as per night or per week.

\subsection{Light curve combination and astrometry}
\label{sec:KS_Lightcurve_Combination}
The meridian flip of the telescope (see \S \ref{sec:KS_Telescope}) causes images taken west of the meridian to be rotated by 180 degrees compared to images taken east of the meridian. It is thus necessary to treat images taken on either side of the meridian as completely separate observations. Because the telescope optics are not exactly axisymmetric, the PSFs of stars in the corners of the field of view are not exactly the same in each orientation, and different reference images  for the eastern and western orientations were created to account for the non-symmetrical nature of the distortions. Other factors like flat fielding errors and detector defects also contribute to images obtained in either orientation being slightly different. The data reduction pipeline produces two versions of a light curve for each identified object in each orientation. The first is the raw flux converted, 3$\sigma$ clipped light curve that removes all outlier data points. We call this the 'processed light curve'. The second version of the light curve is a trend filtered light curve. A 90 day median smoothing is applied to remove long term trends before we use the Trend Filtering Algorithm (TFA) \citep{Kovacs:2005} to remove common systematics in the light curves. We use the closest 150 neighbouring stars that are within three instrumental magnitudes as an input template to detrend the light curve for each star. This detrended light curve is called the 'TFA light curve'.

We use the \textsc{astrometry.net} package \citep{Lang:2010} to find astrometric solutions for each of our reference images (east and west reference image separately). \textsc{astrometry.net} performs astrometric image calibrations without any prior input aside from the data in the image pixels themselves. The calibration information that \textsc{astrometry.net} provides includes image pointing, orientation, plate scale and full coordinate solutions.

To get a combined light curve for each star, KELT-South objects from the eastern and western reference images are first matched to objects listed in the Tycho-2 catalogue \citep{Hog:2000} separately. KELT-South objects are matched to Tycho-2 objects if the sky projected distance between them is less than 76\arcsec ($\sim$3.5 pixels). We chose this distance as it produced the fewest number of double matches (i.e., when more than one KELT-South object is matched to the same Tycho-2 object), while still producing a relatively large fraction of correctly matched stars. If more than one match was found for a Tycho-2 star in the reference image, we selected the closest KELT-South object as the matching object (unless the magnitude difference indicated a wrong match). Using the Tycho-2 positional information for each KELT-South object, we match the Tycho-2 objects to objects in the UCAC4 catalogue \citep{Zacharias:2013}. Finally the eastern and western median-subtracted light curves (matched using the Tycho-2 IDs) are combined and the median magnitude is used as the final KELT-South instrumental magnitude. 

At the end of the data reduction process, we thus have a final total of two light curves for each object identified, one combined processed light curve and one combined TFA light curve.

\subsection{Candidate selection}
\label{sec:KS_Candidate_Selection}
We present a summary of the KELT-South candidate selection process, which involves a number of procedures, and emphasize the modifications from the steps followed in the discovery paper of KELT-1b \citep{Siverd:2012} from the KELT-North survey.
The matching of the object to Tycho-2 and UCAC4 simultaneously provides proper motions and \textit{B}, \textit{V}, \textit{J}, \textit{H} and \textit{K} apparent magnitudes. Next, giant stars are identified and excluded from our sample using a reduced proper motion cut \citep{Gould:2003} following the procedure outlined by \citet{Collier:2007}, with the slight modifications shown in equations 1 and 2 in \S2.3 of \citet{Siverd:2012}.

For most KELT-South fields, the combined light curves use the eastern orientation ID; however, the KELT-South field 27 (KELT-10b's location) and 28 combined light curves were the first fields processed and were given a unique internal ID in a process similar to that of KELT-North. Using the box-least-squares (BLS) algorithm \citep{Kovacs:2002}, we search the combined light curves for periodic transit signals. The selection cuts are performed on five of the statistics output by the \textsc{vartools} \citep{Hartman:2012} implementation of the BLS algorithm. The selection criteria we use are the same as presented in \citet{Siverd:2012} and are shown in \autoref{tbl:BLS_Selection_Criteria} \citep{Kovacs:2002, Burke:2006, Hartman:2008}. Finally, we apply the same stellar density cut as shown in equations 4 and 5 of \S2.3 of \citet{Siverd:2012} but do not apply the \teff\space cut of $<$ 7500 K. All targets that pass the cuts described here are assigned as candidates and a webpage is created containing all the information from the analysis above, along with additional diagnostic tests based on the Lomb-Scargle (LS) \citep{Lomb:1976,Scargle:1982} and analysis-of-variance (AoV) \citep{Schwarzenberg:1989,Devor:2005} algorithms. 

From this analysis, we typically find anywhere from 100 to 1500 candidates per KELT-South field (depending on the proximity of the field to the galactic plane). The KELT-South team individually votes on each candidate in a field and designates each as either a planet or a false positive (such as an eclipsing binary or spurious detection). All candidates with one or more votes for planet are then discussed. From the discussion, the most promising potential planet candidates are sent for follow-up photometric confirmation, reconnaissance spectroscopy, or in very promising cases, both.
\begin{table}
 \centering
 \caption{KELT-South BLS selection criteria for candidate exoplanet light curves and values for KELT-10b.}
 \label{tbl:BLS_Selection_Criteria}
 \begin{tabular}{lll}
    \hline
    \hline
    BLS Statistic &  Selection & Value for \\
    &  Criteria  & KELT-10b\\
    \hline
    Signal detection efficiency & SDE $>$ 7.0 & 18.25\\
    Signal to pink-noise & SPN $>$ 7.0 & 13.40\\
    Transit depth & $\delta <$ 0.05 & 0.0121\\
    $\chi^2$ ratio & $\displaystyle\frac{\Delta\chi^2}{\Delta\chi^2_{-}} >$ 1.5 & 5.39\\
    Duty cycle & q $<$ 0.1 & 0.04\\
    \hline
    \hline
 \end{tabular}
\end{table}  


\begin{figure*}
  \includegraphics[width=\textwidth]{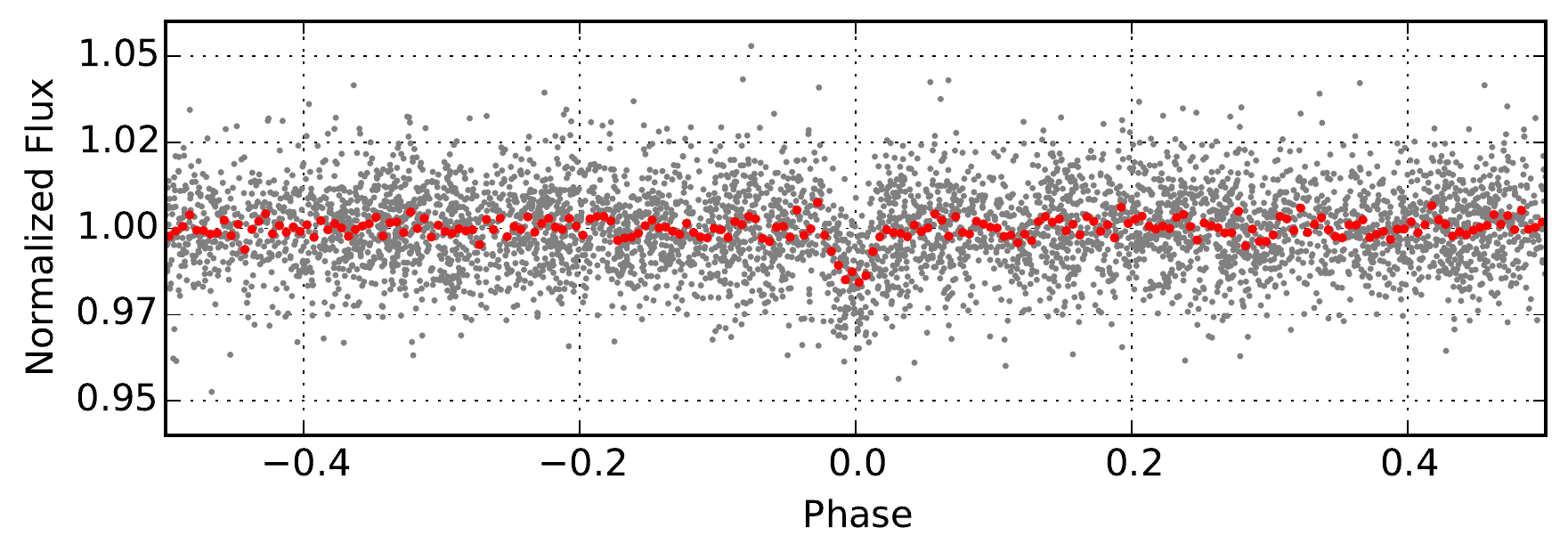}
  \caption{Discovery light curve of KELT-10b from the KELT-South telescope. The light curve contains 4967 observations spanning just over 4 years, phase folded to the orbital period of P = 4.1664439 days. The red points show the light curve binned in phase using a bin size of 0.005 ($\approx$ 30 minutes). These points are shown for illustrative purposes only and are not used in the final global fit described in \autoref{sec:Global_Modeling}. (A colour version of this figure is available in the online journal. A table containing all the KELT-South measurements is available in a machine-readable form in the online journal.)}
  \label{fig:KS_Discovery_Lightcurve}
\end{figure*}

\section{Discovery and follow-up observations}
\label{sec:Discovery_Follow_Up}

\subsection{KELT-South observations and photometry}
\label{sec:KS_Photometry}

KELT-10b is located in the KELT-South field 27 with field centre located at J2000 $\alpha$ = 19$^{h}$55$^{m}$48$^{s}$ and $\delta$=-53\degr00\arcmin00\arcsec. Field 27 was monitored from UT 2010 March 19 to UT 2014 April 18 with over 5000 images acquired during that time (4967 observations were used in the final light curve). We reduced the raw survey data using the pipeline described in \S\ref{sec:KS_Survey} and performed candidate vetting and selection as described in \S\ref{sec:KS_Candidate_Selection}. One of the candidates from field 27 (KS27C013526) was star TYC 8378-64-1 / 2MASS 18581160-4700116, located at J2000 $\alpha$=18$^{h}$58$^{m}$11$\fs$601 and $\delta$=-47\degr00\arcmin11$\farcs$68. The star has Tycho magnitudes B$_T$ = 11.504 and V$_T$ = 10.698 \citep{Hog:2000}.  Full catalog properties of this star are provided in \autoref{tbl:Host_Lit_Props}.  The light curve of KS27C013526 has a weighted RMS of 11.9 milli-magnitudes (mmag), which is typical for KELT-South light curves with magnitude around \textit{V} = 11. A significant BLS signal was found at a period of P $\simeq$ 4.1664439 days, with a transit depth of $\delta$ $\simeq$ 12.1 mmag. Further detection statistics are listed in \autoref{tbl:BLS_Selection_Criteria}. The discovery light curve of KELT-10b is shown in \autoref{fig:KS_Discovery_Lightcurve}.

\begin{table*}
 \centering
 \caption{Stellar Properties of KELT-10 obtained from the literature.}
 \label{tbl:Host_Lit_Props}
 \begin{tabular}{llccl}
 \hline
 \hline
  Parameter & Description & Value & Source & Reference(s) \\
 \hline
Names 			& 					& TYC 8378-64-1 		& 		&			\\
			& 					& 2MASS 18581160-4700116 	& 		&			\\
			& 					& GSC 08378-00064 		& 		&			\\
			&					&				&		&			\\
$\alpha_{J2000}$	& Right Ascension (RA)			& 18:58:11.601			& Tycho-2	& \citet{Hog:2000}	\\
$\delta_{J2000}$	& Declination (Dec)			& -47:00:11.68			& Tycho-2	& \citet{Hog:2000}	\\
B$_T$			& Tycho B$_T$ magnitude			& 11.504 $\pm$ 0.078		& Tycho-2	& \citet{Hog:2000}	\\
V$_T$			& Tycho V$_T$ magnitude			& 10.698 $\pm$ 0.061		& Tycho-2	& \citet{Hog:2000}	\\
			&					&				&		&			\\
Johnson V		& APASS magnitude			& 10.854 $\pm$ 0.043		& APASS 	& \cite{Henden:2015}	\\
Johnson B		& APASS magnitude			& 11.462 $\pm$ 0.006		& APASS 	& \cite{Henden:2015}	\\
Sloan g'		& APASS magnitude			& 11.107 $\pm$ 0.011		& APASS 	& \cite{Henden:2015}	\\
Sloan r'		& APASS magnitude			& 10.666 $\pm$ 0.012		& APASS 	& \cite{Henden:2015}	\\
Sloan i'		& APASS magnitude			& 10.730 $\pm$ 0.123		& APASS 	& \cite{Henden:2015}	\\
			&					&				&		&			\\
J			& 2MASS magnitude			& 9.693 $\pm$ 0.019		& 2MASS 	& \citet{Cutri:2003, Skrutskie:2006}	\\
H			& 2MASS magnitude			& 9.388 $\pm$ 0.022		& 2MASS 	& \citet{Cutri:2003, Skrutskie:2006}	\\
K			& 2MASS magnitude			& 9.338 $\pm$ 0.021		& 2MASS 	& \citet{Cutri:2003, Skrutskie:2006}	\\
			&					&				&		&			\\
\textit{WISE1}		& WISE passband				& 9.305 $\pm$ 0.022		& WISE 		& \citet{Wright:2010, Cutri:2014}	\\
\textit{WISE2}		& WISE passband				& 9.350 $\pm$ 0.019		& WISE 		& \citet{Wright:2010, Cutri:2014}	\\
\textit{WISE3}		& WISE passband				& 9.275 $\pm$ 0.034		& WISE 		& \citet{Wright:2010, Cutri:2014}	\\
			&					&				&		&			\\
$\mu_{\alpha}$		& Proper Motion in RA (mas yr$^{-1}$)	& 0.4 $\pm$ 1.0 		& UCAC4		& \cite{Zacharias:2013} \\
$\mu_{\delta}$		& Proper Motion in DEC (mas yr$^{-1}$)	& -13.2 $\pm$ 1.3		& UCAC4		& \cite{Zacharias:2013} \\
			&					&				&		&			\\
U			& Calculated space motion (\kms)	& 34.6 $\pm$ 0.5		& This work	& \\
V			& Calculated space motion (\kms)	& -2.3 $\pm$ 1.3		& This work	& \\
W			& Calculated space motion (\kms)	& -8.4 $\pm$ 0.9		& This work	& \\
$d$			& Distance (pc) 			& 183 $\pm$ 14			& This work	& \\
$A_V$			& Visual Extinction (magnitude)		& $0.17^{+0.00}_{-0.16}$	& This work	& \\
			& Age (Gyr)				& 4.5$\pm$0.7			& This work	& \\
\hline
\hline
\end{tabular}
\end{table*}

\subsection{Follow-up photometry}
\label{sec:Follow-up_Photometry}

\begin{figure}
\includegraphics[width=\columnwidth]{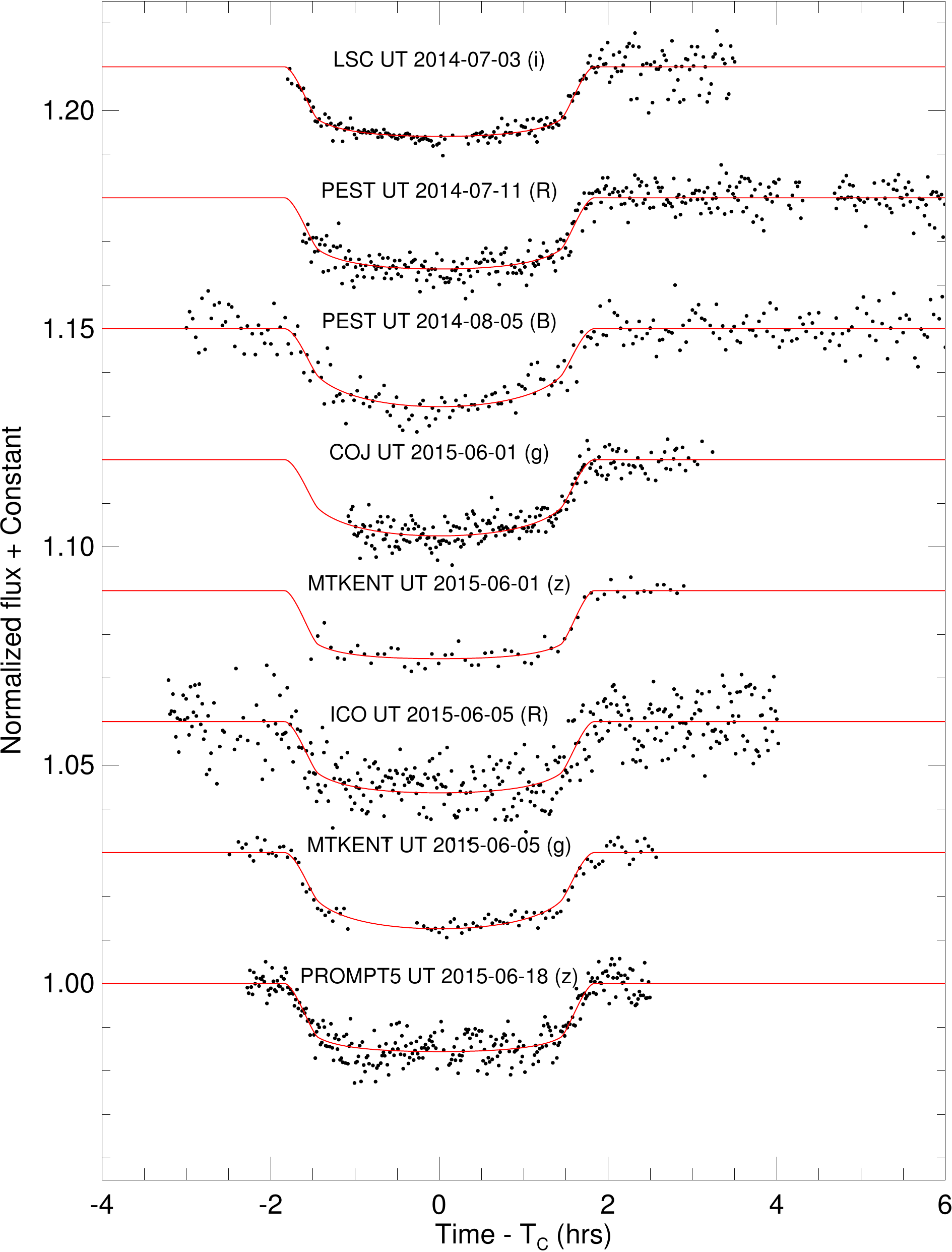}
\includegraphics[width=\columnwidth]{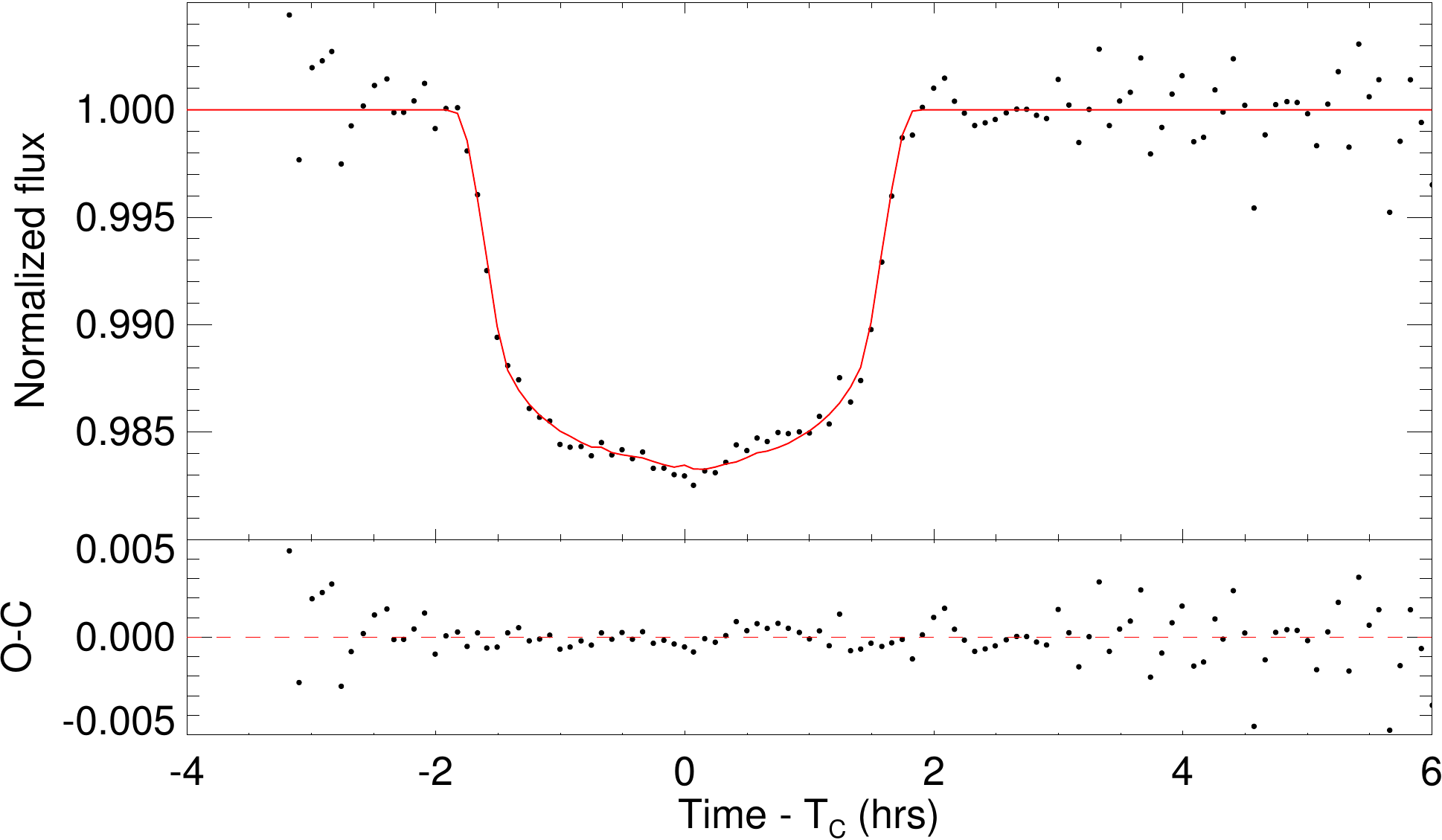}
\caption{\textit{Top}: Follow-up transit photometry of KELT-10. The red line is the best-fit transit model (discussed in \S\ref{sec:Global_Modeling}). The labels are: LSC = CTIO LCOGT observations (see \S\ref{sec:LCOGT_Obs}, PEST = Perth Exoplanet Survey Telescope observations (see \S\ref{sec:PEST_Obs}), COJ = SSO LCOGT observations (see \S\ref{sec:LCOGT_Obs}), ICO = Adelaide Observatory (Ivan Curtis Observatory) observations (see \S\ref{sec:Adelaide_Obs}), MTKENT = Mt. Kent Observatory observations (see \S\ref{sec:MtKent_Obs}), and Prompt5 = CTIO Skynet observations (see \S\ref{sec:Skynet_Obs}). \textit{Bottom}: All the follow-up light curves combined and binned in 5 minute intervals. This light curve is not used for analysis, but rather to show the best combined behaviour of the transit. The solid red line shows the transit model from the global fitting procedure described in \S\ref{sec:Global_Modeling} for each of the individual light curves. (A colour version of this figure is available in the online journal.)}
\label{fig:All_Lightcurve}
\end{figure}

\begin{table*}
 \centering
 \caption{Photometric follow-up observations and the detrending parameters found by AIJ for the global fit.}
 \label{tbl:detrending_parameters}
 \begin{tabular}{llll}
    \hline
    \hline
    Follow-up Observations & Date (UT) & Filter & Detrending parameters used in the global fit \\
    \hline
    LCOGT & 2014 July 3 & $i^{\prime}$ & airmass \\
    PEST & 2014 July 11 & \textit{R} & time, FWHM, \textit{xy} coordinates \\
    PEST & 2014 August 5 & \textit{B} & sky counts, \textit{xy} coordinates \\
    LCOGT & 2015 June 1 & $g^{\prime}$ & airmass, total counts \\
    Adelaide & 2015 June 1 & \textit{R} & airmass, sky counts, total counts \\
    Mt. Kent & 2015 June 1 & $z^{\prime}$ & airmass, FWHM, sky counts \\
    Mt. Kent & 2015 June 5 & $g^{\prime}$ & airmass, total counts \\
    Skynet & 2015 June 18 & $z^{\prime}$ & airmass, meridian flip, FWHM, total counts \\
    \hline
    \hline
 \end{tabular}
\begin{flushleft}
  \footnotesize \textbf{\textsc{NOTES}} \\
  \footnotesize The tables for all the follow-up photometry are available in machine-readable form in the online journal.
  \end{flushleft}
\end{table*}  

We obtained follow-up time-series photometry of KELT-10b to check for false positives, better determine the transit shape, and better determine the phase and period of the transiting exoplanet. We used the \textsc{TAPIR} software package \citep{Jensen:2013} to predict transit events, and we obtained 8 full or partial transits in multiple bands between 2014 July and 2015 June. All data were calibrated and processed using the AstroImageJ package (AIJ)\footnote{http://www.astro.louisville.edu/software/astroimagej} \citep{Collins:2013, Collins:2016} unless otherwise stated. For all follow-up photometric observations, we use the AIJ package to determine the best detrending parameters to include in the global fit. \autoref{tbl:detrending_parameters} lists the follow-up photometric observations and the detrending parameters determined by AIJ. All follow-up photometry is shown in \autoref{fig:All_Lightcurve}.  

\subsubsection{LCOGT observations}
\label{sec:LCOGT_Obs}

The Las Cumbres Observatory Global Telescope (LCOGT) network\footnote{http://lcogt.net/} consists of a globally distributed set of telescopes. We observed a nearly full transit of KELT-10b in the Sloan $i^\prime$-band on UT 2014 July 03 from a LCOGT 1 m telescope at Cerro Tololo Inter-American Observatory (CTIO), with a 4K\space$\times$\space4K Sinistro detector with a 27\arcmin\space$\times$\space27\arcmin\space field of view and a pixel scale of 0\farcs39 pixel$^{-1}$, labelled as 'LSC' in \autoref{fig:All_Lightcurve}).  We also observed a partial transit in the Sloan $g^\prime$-band on UT 2015 June 01 from a 1 m LCOGT telescope at Siding Springs Observatory (SSO, labelled as 'COJ' in \autoref{fig:All_Lightcurve}, with a 4K\space$\times$\space4K SBIG Science camera with a 16$^\prime \times$16$^\prime$ field of view and a pixel scale of 0\farcs23 pixel$^{-1}$. The reduced data were downloaded from the LCOGT archive and observations were analysed using custom routines written in GDL.\footnote{GNU Data Language; http://gnudatalanguage.sourceforge.net/}. 

\subsubsection{Skynet observations}
\label{sec:Skynet_Obs}

The Skynet network \citep{Reichart:2005}\footnote{https://skynet.unc.edu/} is another set of distributed telescopes.  We observed a partial transit of KELT-10b in the Sloan $z^\prime$-band on UT 2015 June 18 from CTIO using the 0.4 m Prompt5 telescope from the PROMPT (Panchromatic Robotic Optical Monitoring and Polarimetry Telescope) subset of Skynet telescopes.  The Prompt5 telescope hosts an Alta U47+s camera by Apogee, which has an E2V 1024$\times$1024 CCD, a field of view of 10\arcmin\space$\times$\space10\arcmin, and a pixel scale of 0\farcs59 pixel$^{-1}$. Reduced data were downloaded from the Skynet website and analysed using custom GDL routines.

\subsubsection{PEST observations}
\label{sec:PEST_Obs}

The PEST (Perth Exoplanet Survey Telescope) observatory is a backyard observatory owned and operated by Thiam-Guan (TG) Tan. It is equipped with a 12 inch Meade LX200 SCT f/10 telescope with focal reducer yielding f/5. The camera is an SBIG ST-8XME, and focusing is computer controlled with an Optec TCF-Si focuser. CCD Commander is used to script observations, FocusMax for focuser control, and CCDSoft for control of the CCD camera. PinPoint is used for plate solving.  A partial transit (covering the last half of ingress to almost 2 hours after egress) was observed in \textit{R} on UT 2014 July 11. A full transit in \textit{B} was observed on UT 2014 August 5.

\subsubsection{Adelaide observations}
\label{sec:Adelaide_Obs}

The Adelaide Observatory owned and operated by Ivan Curtis is largely run manually from his back yard in Adelaide, South Australia. The observatory is equipped with a 9.25 inch Celestron SCT telescope with an Antares 0.63$\times$ focal reducer yielding an overall focal ratio of f/6.3. The camera is an Atik 320e, which uses a cooled Sony ICX274 CCD of 1620\space$\times$\space1220 pixels. The field of view is $\sim$ 16\farcm6 $\times$ 12\farcm3 with a resolution of 0\farcs62 pixel$^{-1}$.  We observed a complete transit of KELT-10b in \textit{R} on UT 2015 June 5 (see light curve 'ICO' in \autoref{fig:All_Lightcurve}).

\subsubsection{Mt. Kent observations}
\label{sec:MtKent_Obs}
A partial transit of KELT-10b was observed in the Sloan $z^\prime$-band on UT 2015 June 1 and a full transit was observed in the Sloan $g^\prime$-band on UT 2015 June 5 (see \autoref{fig:All_Lightcurve}) from the Mt. Kent Observatory, located outside Toowoomba, Queensland, Australia. All observations were performed with a defocused 0.5 m Planewave CDK20 telescope equipped with an Alta U16M Apogee camera. This combination produces a 36\farcm7 $\times$ 36\farcm7 field of view and a pixel scale of 0\farcs5375 pixel$^{-1}$.  Conditions on both nights were clear, however both observations were taken close to full Moon.

\subsection{Spectroscopic follow-up}
\label{sec:Spec_Follow-up}

\subsubsection{Spectroscopic reconnaissance}
\label{sec:recon}

In order to rule out eclipsing binaries and other common causes of false positives for transiting exoplanets surveys, we acquire spectra of KELT-South candidates using the Wide Field Spectrograph (WiFeS) \citep{Dopita:2007} on the Australian National University (ANU) 2.3~m telescope at Siding Spring Observatory in Australia. We follow the observing methodology and data reduction pipeline as set out in \citet{Bayliss:2013}. A single 100 s spectrum of KELT-10 was obtained using the B3000 grating of WiFeS on 2014 July 9, which delivers a spectrum of R~=~3000 from 3500\AA to 6000\AA. From this spectrum we determined KELT-10 to be a dwarf star with an effective temperature of $6100\pm100$ K. We then obtained four higher resolution spectra (with exposure times of $\sim$ 60 s), spanning the nights 2014 July 9 to 2014 July 11, with the R7000 grating of WiFeS. This grating delivers a spectral resolution of R~=~7000 across the wavelength region 5200 to 7000 \AA, and provides radial velocities precise to the $\pm$1~\kms\space level. From these four measurements we determined that KELT-10 did not display any radial velocity variations above the level of K~=~1~\kms, ruling out eclipsing binaries with high amplitude radial velocity variations. The candidate was therefore deemed to be a high priority target for high precision radial velocity follow-up (see \autoref{sec:rv}).

To date we have vetted 105 KELT-South candidates using the WiFeS on the ANU 2.3 m telescope, of which 67 (62.6\%) have been ruled out as eclipsing binaries due to high-amplitude, in-phase, radial velocity variations.  A further two candidates (1.9\%) were ruled out after they were identified as giants. Typically each KELT-South candidate requires only 5 minutes of \textit{total} exposure time for this vetting, making it an extremely efficient method of screening candidates compared to photometric follow-up or high-resolution spectroscopy.

\subsubsection{High precision spectroscopic follow-up}
\label{sec:rv}

\begin{figure}
  \centering \includegraphics[width=\columnwidth]{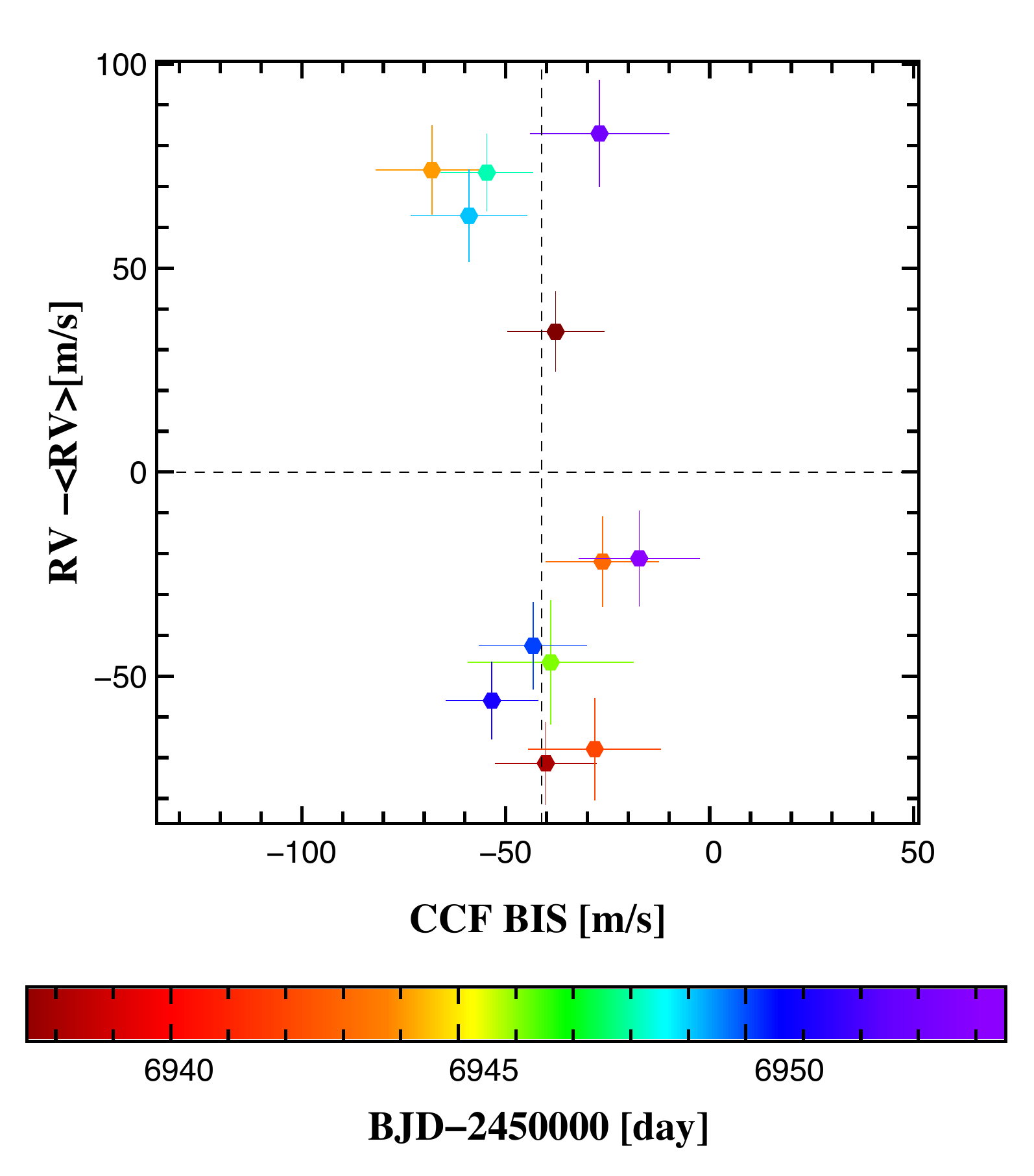}
  \caption{Bisector measurements for the CORALIE spectra used for radial velocity measurements. We find no significant correlation between RV and the bisector spans. (A colour version of this figure is available in the online journal.)}
  \label{fig:RV_BS}
\end{figure}

Multi-epoch, high resolution spectroscopy with a stabilized spectrograph can provide high precision radial velocity measurements capable of confirming the planetary nature of a transiting exoplanet candidate and precisely measuring its true mass. We obtained high-precision radial velocity measurements for KELT-10 using the CORALIE spectrograph \citep{Queloz:2001} on the Swiss 1.2 m Leonard Euler telescope at the ESO La Silla Observatory in Chile. CORALIE is a fibre-fed echelle spectrograph which provides stable spectra with spectral resolution of R = 60000. Spectra are taken with a simultaneous exposure of a Thorium-Argon discharge lamp acquired through a separate calibration fibre, which allows for radial velocities to be measured at levels better than 3~\ms\space for bright stars \citep{Pepe:2002}.
Data are reduced and radial velocities computed in real time via the standard CORALIE pipeline, which cross-correlates the stellar spectra with a numerical mask with non-zero zones corresponding to stellar absorption features at zero velocity.  
Radial velocities were measured for KELT-10 over 12 different epochs between 2014 October 7 and 2014 October 23, spanning a range of orbital phases. The results are presented in \autoref{tbl:CORALIE_RV}. The radial velocities displayed a sinusoidal variation with a period and phase that matched the orbit as determined from the discovery and follow-up photometry. The peak-to-peak amplitude of the variation was 154~\ms, indicative of a short-period hot Jupiter.  Bisector spans were measured using the CCF peak as described in \citep{Queloz:2001b} and no correlation is evident between the bisector spans and the radial velocities (see \autoref{fig:RV_BS}), helping rule out a blended eclipsing binary system \citep{Gray:1983,Gray:2005}.  These radial velocities are used in the global modelling of the KELT-10 system as set out in \autoref{sec:Global_Modeling}.

\label{sec:Spectroscopy}
\begin{table}
 \centering
 \caption{KELT-10 radial velocity observations with CORALIE.}
 \label{tbl:CORALIE_RV}
 \begin{tabular}{ccc}
 \hline
 \hline
  $\bjdtdb$ & RV & RV error\\
   &(\ms)&  (\ms) \\
 \hline
  2456937.517729 & 31942.72 & 9.73\\
  2456938.541050 & 31836.89 & 10.06\\
  2456942.492638 & 31840.37 & 12.48\\
  2456943.496229 & 31886.32 & 10.97\\
  2456944.494299 & 31982.29 & 10.83\\
  2456946.495001 & 31861.65 & 15.14\\
  2456948.521337 & 31981.69 & 9.40\\
  2456949.497177 & 31971.14 & 11.19\\
  2456950.523627 & 31865.72 & 10.55\\
  2456951.521526 & 31852.24 & 9.39\\
  2456953.505857 & 31991.27 & 12.97\\
  2456954.499325 & 31887.12 & 11.57\\
 \hline
 \hline
\end{tabular}
\begin{flushleft}
  \footnotesize \textbf{\textsc{NOTES}} \\
  \footnotesize This table is available in its entirety in a machine-readable form in the online journal.
  \end{flushleft}
\end{table}

\begin{figure}
  \centering \includegraphics[width=\columnwidth]{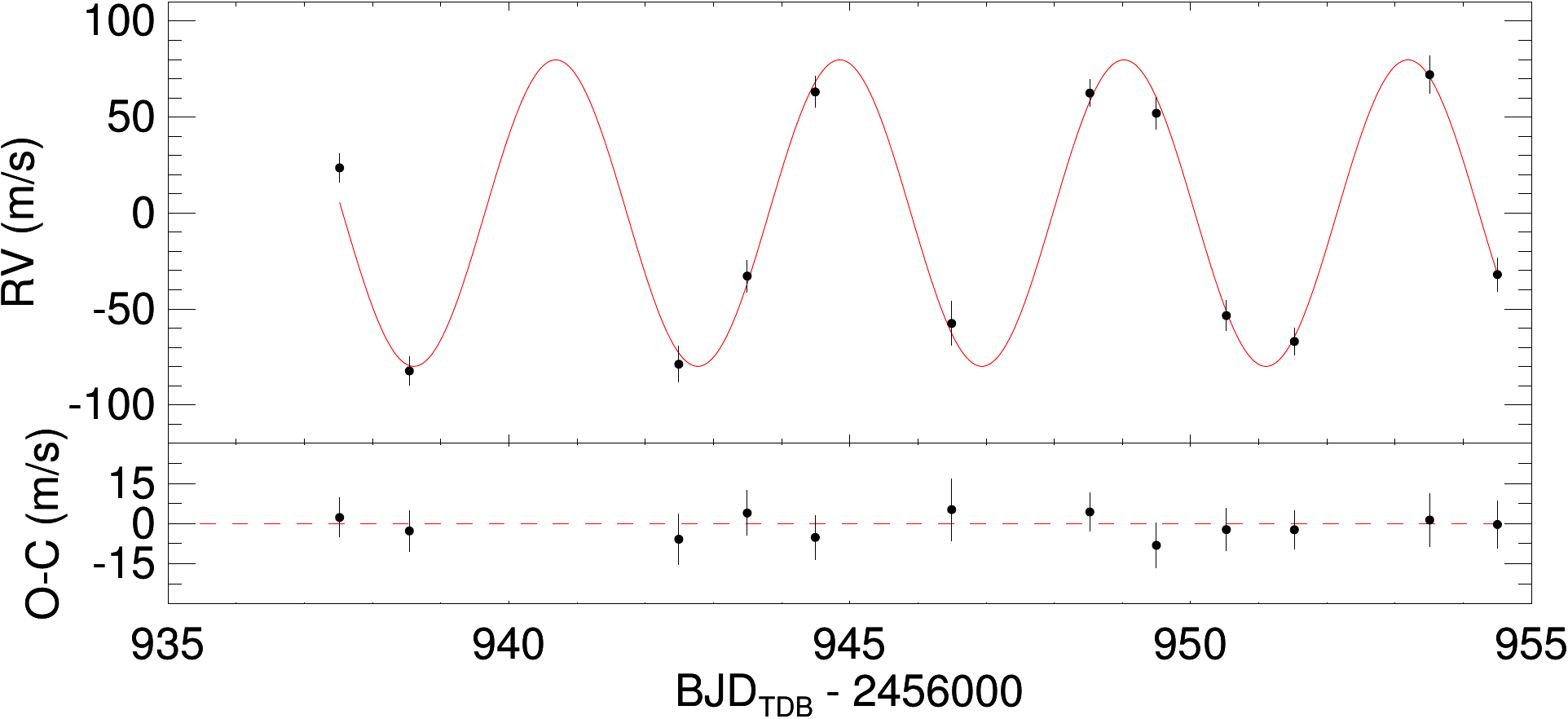}
  \caption{CORALIE radial velocity measurements and residuals for KELT-10. The best fitting model is shown in red. The bottom panel shows the residuals of the RV measurements to the best fitting model. (A colour version of this figure is available in the online journal.)}
  \label{fig:RV_Phased}
\end{figure}

\begin{figure}
  \centering \includegraphics[width=\columnwidth]{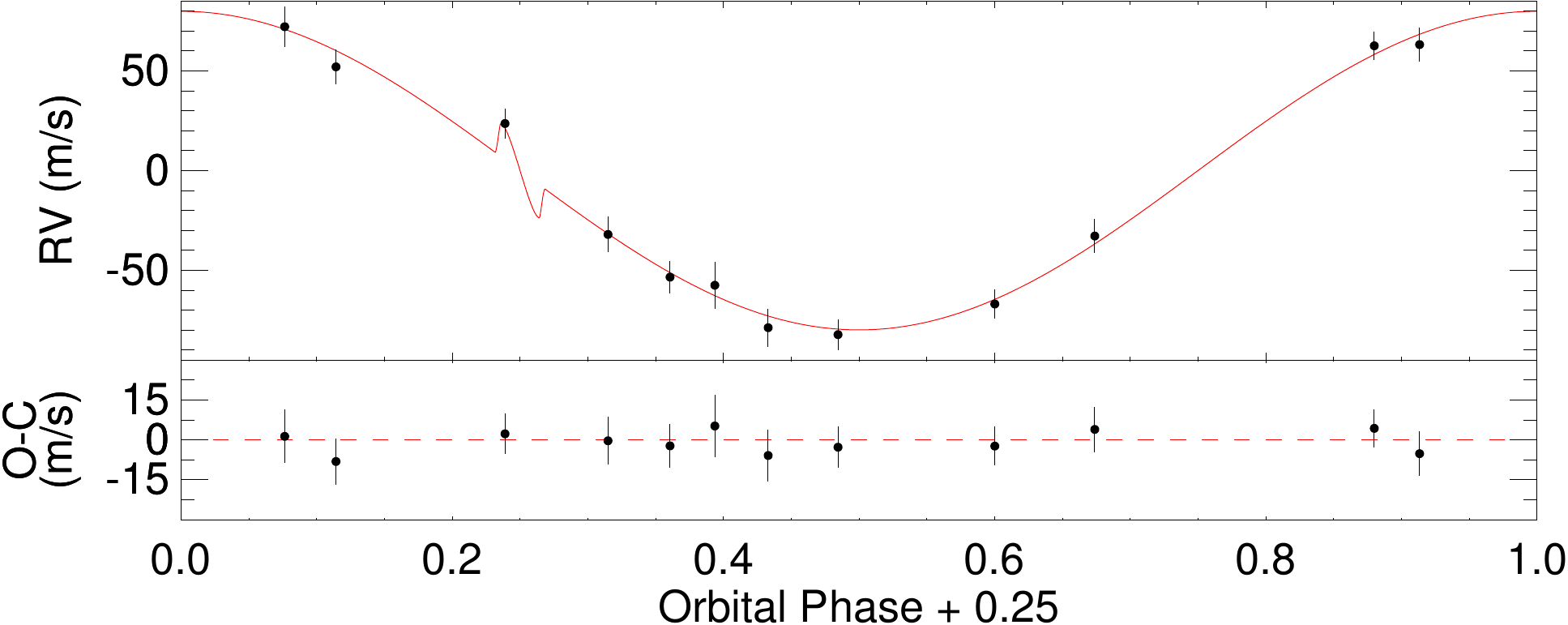}
  \caption{Radial velocity measurements phase-folded to the best fitting linear ephemeris. The best fitting model is shown in red. The predicted Rossiter-McLaughlin effect assumes perfect spin-orbit alignment and it is not well constrained by our data. The bottom panel shows the residuals of the RV measurements to the best fitting model. (A colour version of this figure is available in the online journal.)}
  \label{fig:RV_Unphased}
\end{figure}

\subsection{Adaptive optics observations}
\label{sec:AO_Obs}
\begin{figure}
  \centering \includegraphics[width=\columnwidth]{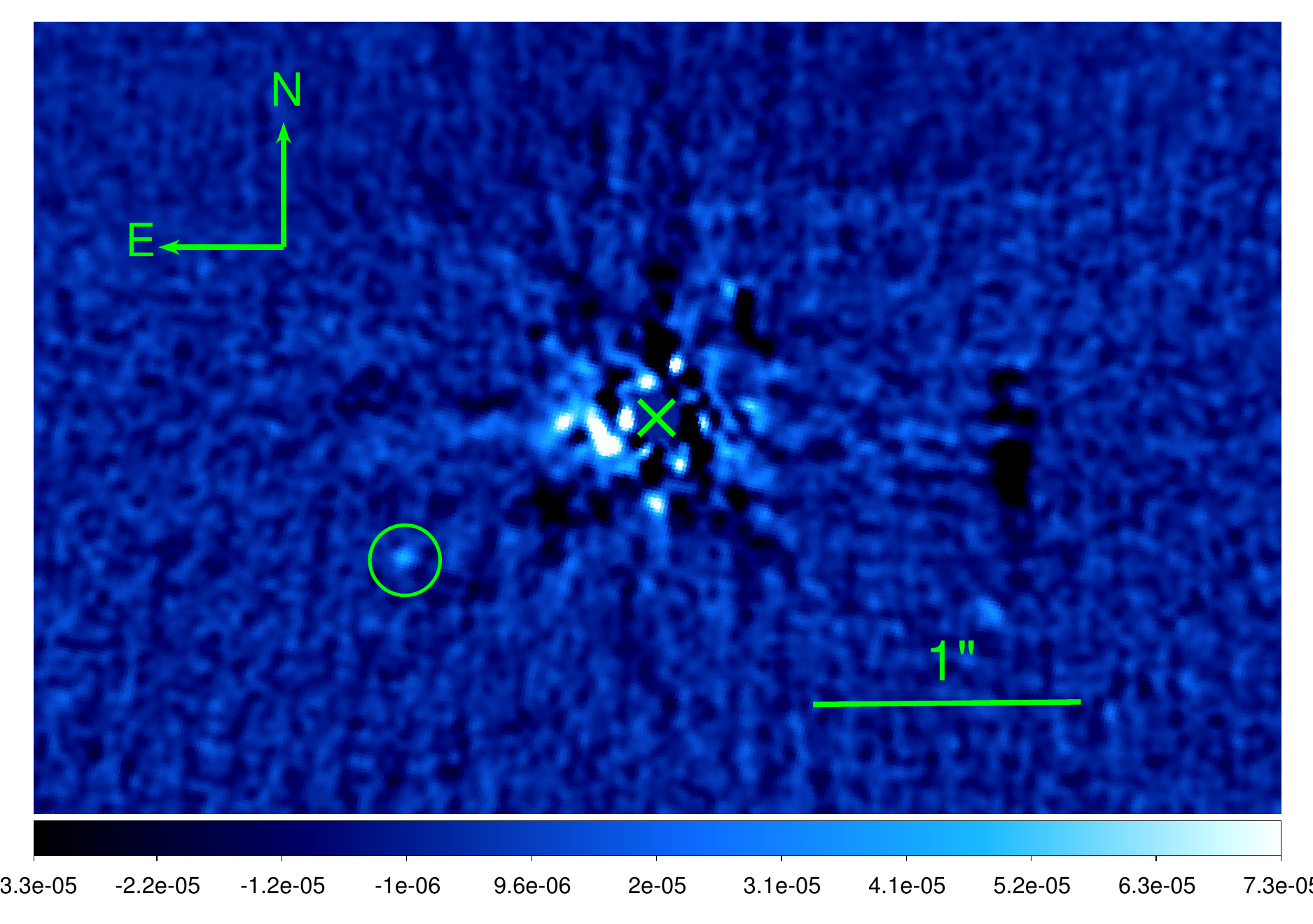}
  \caption{The NAOS-CONICA adaptive optics image of KELT-10. The location of KELT-10 is shown by the green cross. North is up, east is left. A horizontal 1\arcsec\space bar is also shown for scale. A faint companion (circled in green) with $\Delta$K~=~$9\pm0.3$ mag located 1$\farcs$1 to the SE of KELT-10 is clearly visible. (A colour version of this figure is available in the online journal.)}
  \label{fig:AO_Obs}
\end{figure}
Adaptive optics (AO) imaging places limits on the existence of nearby eclipsing binaries that could be blended with the primary star KELT-10 at the resolution of the KELT-South and follow-up data, thereby causing a false-positive planet detection. In addition, it places limits on any nearby blended source that could contribute to the total flux, and thereby result in an underestimate of the transit depth and thus planet radius in the global fit presented in \autoref{sec:Global_Modeling}.

We observed KELT-10 with the Nasmyth Adaptive Optics System (NAOS) Near-Infrared Imager and Spectrograph (CONICA) instrument \citep{Lenzen:2003,Rousset:2003} on the Very Large Telescope (VLT) located at the ESO Paranal observatory as part of the program 095.C-0272(A) 'Adaptive Optics imaging of the Brightest Transiting Exoplanets from KELT-South' (PI: Mawet) on UT 2015 August 7. We used the Ks-band filter and the S13 camera which has a plate scale of 0$\farcs$013 pixel$^{-1}$.

The images were acquired as a sequence of 30 dithered exposures. Each exposure was the average of 10 frames with 6 second integration time, making for a total open shutter time of 1800 seconds. We used the pupil tracking mode where the instrument co-rotates with the telescope pupil to fix diffraction and speckles to the detector reference frame, allowing the sky to counter-rotate with the parallactic angle, effectively enabling angular differential imaging.

We reduced the data by subtracting a background image made out of median-combined dithered frames, dividing by a flat field and interpolating for bad pixels and other cosmetics. The reduced images were then processed using principal component analysis \citep{Soummer:2012} to mitigate speckle noise, yielding the image shown in \autoref{fig:AO_Obs}. 

\begin{figure}
  \centering \includegraphics[width=\columnwidth]{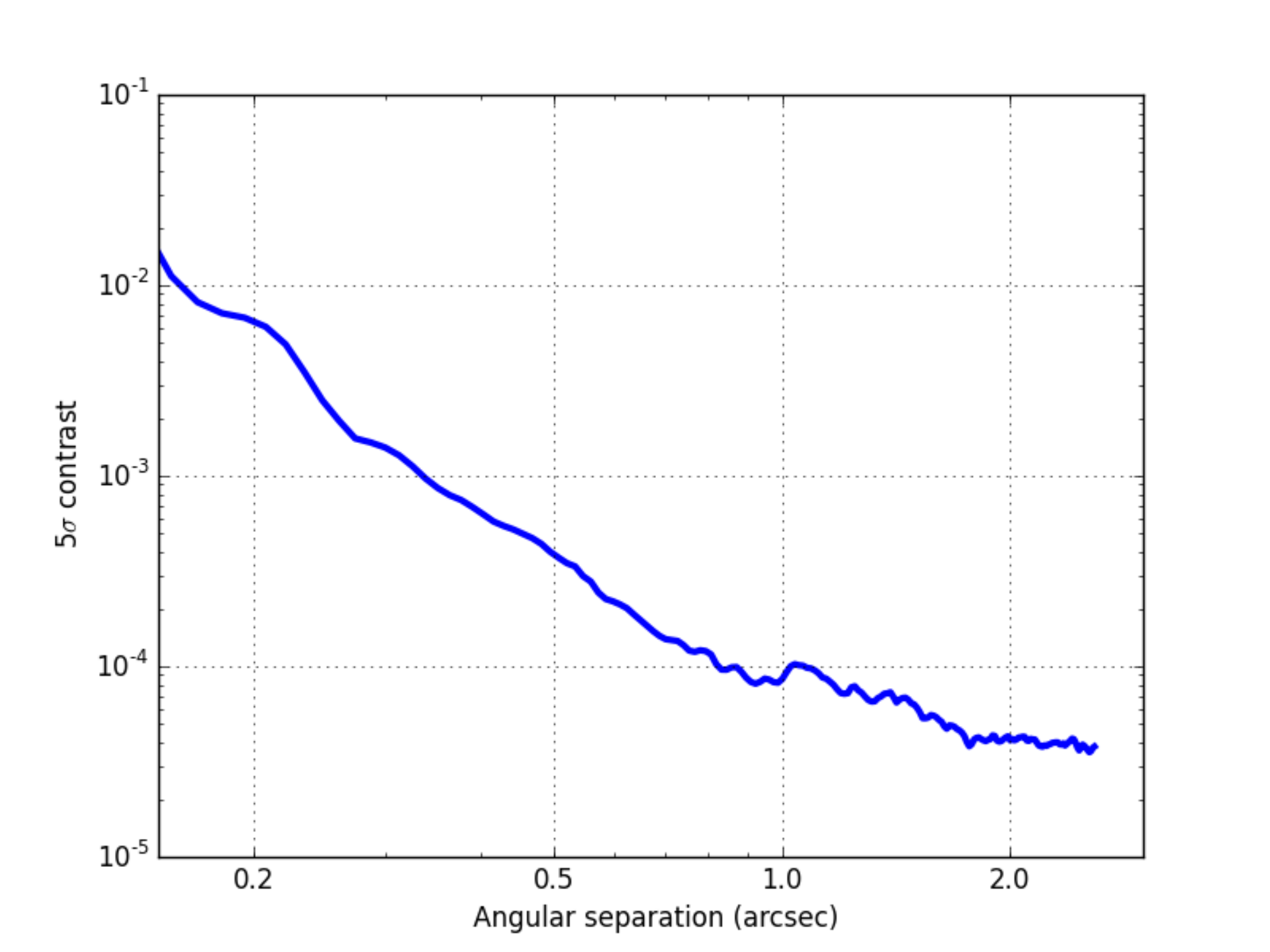}
  \caption{5$\sigma$ detection limit as a function of angular separation derived from the NAOS-CONICA adaptive optics image around KELT-10. (A colour version of this figure is available in the online journal.)}
  \label{fig:AO_concurve}
\end{figure}

We calculate the 5$\sigma$ contrast limit as a function of angular separation by defining a series of concentric annuli centred on the star. For each of the concentric annuli, we calculated the median and the standard deviation of flux for pixels within this annulus. We used the value of five times the standard deviation above the median as the 5$\sigma$ detection limit. The resulting 5$\sigma$ contrast curve is presented in \autoref{fig:AO_concurve}.

An off-axis point source with a contrast of $\Delta$K~=~$9\pm0.3$ magnitude to the primary is detected at a distance of 1$\farcs$1$\pm$0$\farcs$013 at position angle of $\simeq 119^\circ$.  At a separation of 1$\farcs$1, it is unclear whether or not the companion is bound to KELT-10. If the companion is bound, we calculate an absolute M$_K$ = 12.026 at a distance of 183$\pm$14 pc. Using the Baraffe \citep{Baraffe:2002} models we estimate a mass of 0.073 $M_{\sun}$ for the companion, which is consistent with a very late-M dwarf star, just barely above the sub-stellar boundary.

Additional high-contrast follow-up imaging of the system could help determine whether or not the faint companion is bound, but the proper motion of KELT-10 is not very large. If the companion were a distant background source (thus $\sim$stationary), it would separate from KELT-10 with a speed comparable to proper motion of KELT-10. A 15-year baseline in follow-up imaging would only provide a movement of 270 milli-arcsec, which necessitates highly accurate astrometric and long term monitoring to determine the physical association between the two stars. 

To test the probability of chance alignment for the KELT-10 binary system, we used an statistical model of the Galactic stellar distribution \citep{Dhital:2010}. A more detailed description of this analysis is shown in \citet{Dhital:2010} but we provide a brief, pertinent description here. The Galactic model is parameterized by an empirically measured stellar number density distribution in a 30\arcmin $\times$ 30\arcmin conical volume centered at Galactic position of KELT-10. The number density distributions are constrained by empirical measurements from the Sloan Digital Sky Survey \citep{Juri:2008, Bochanski:2010} and accurately account for the decrease in stellar number density with both Galactocentric radius and Galactic height. Using the Galactic model to calculate the frequency of unrelated, random pairings within the volume defined by the KELT-10 binary, assuming that all the simulated stars are, by definition, single. Thus, any random pairing was flagged as a chance alignment. In 10$^6$ independent simulations, only 86 produced a random pairing, equating to a chance alignment probability of 0.0086\%. This provides strong evidence that KELT 10B is a binary companion to KELT-10A (TYC 8378-64-1).

\section{Host star properties}
\label{sec:Host_Props}


\subsection{SME stellar analysis}
\label{sec:SME_Analysis}
In order to determine precise stellar parameters for KELT-10, we utilized the 12 high-resolution R $\sim$ 55,000, low S/N $\sim$ 10--20 CORALIE spectra acquired for radial velocity confirmation of KELT-10b. For each of the individual spectra, we continuum normalized each echelle order, stitched the orders into a single 1-D spectrum, and shifted the wavelengths to rest velocity by accounting for barycentric motion, the space velocity of KELT-10, as well as the radial velocity induced by KELT-10b. The 12 individual 1-D spectra were then co-added using the median of each wavelength resulting in a single spectrum with a S/N $\sim$ 30--50, sufficient for detailed spectroscopic analysis.

Stellar parameters for KELT-10 are determined using an implementation of Spectroscopy Made Easy (SME) \citep{Valenti:1996}. We base the general method of our SME analysis on that given in \citet{Valenti:2005}; however, we use a line list, synthesized wavelength ranges, and abundance pattern adapted from \citet{Stempels:2007} and \citet{Hebb:2009}, and also incorporate the newest MARCS model atmospheres \citep{Gustafsson:2008}. The details of our technique are outlined in the recent analysis of WASP-13 \citep{Gomez:2013}. Briefly, we have expanded on the technique outlined in \citet{Valenti:2005} that allows us to operate SME in an automated fashion by utilizing massive computational resources. We have developed an extensive Monte Carlo approach to using SME by randomly selecting 500 initial parameter values from a multivariate normal distribution with 5 parameters: effective temperature (\teff), surface gravity ($\log{g_*}$), iron abundance (\feh), metal abundance ([m/H]), and rotational velocity of the star ($v\sin{I_*}$). We then allow SME to find a best fitting synthetic spectrum and solve for the free parameters for the full distribution of initial guesses, producing 500 best-fitting solutions for the stellar parameters. We determine our final measured stellar properties by identifying the output parameters that give the optimal SME solution (i.e., the solution with the lowest $\chi^{2}$). The overall SME measurement uncertainties in the final parameters are calculated by adding in quadrature: 1) the internal error determined from the 68.3\% confidence region in the $\chi^{2}$ map, and 2) the median absolute deviation of the parameters from the 500 output SME solutions to account for the correlation between the initial guess and the final fit. 

Our final SME spectroscopic parameters for KELT-10 are: \teff~=~5925$\pm$79~K (in good agreement with the independently determined \teff\space found by the ANU analysis), $\log{g_*}$~=~4.38$\pm$0.17, [m/H]~=~0.05$\pm$0.06, \feh~=~0.12$\pm$0.11 and a projected rotational velocity $v\sin{I_*}$ $<$ 3.6 \kms. We constrain the macro- and microturbulent velocities to the empirically constrained relationship, similar to \citet{Gomez:2013}; however, we do allow them to change during our modelling according to the other stellar parameters. Our best fitting stellar parameters result in \textit{v}$_{\textup{mac}}$~=~4.14~\kms\space and \textit{v}$_{\textup{mic}}$~=~1.05~\kms. We note the large error bars on \feh\space are likely due to the relatively low S/N of the co-added CORALIE spectrum (estimated to be S/N $\sim$ 30--50). Also, we find that we are unable to resolve the rotational velocity ($v\sin{I_*}$) of KELT-10 from the instrument profile of CORALIE, likely due to the combination of the low S/N of the co-added spectrum and the star's slow rotation. We therefore are only able to infer an upper limit of $v\sin{I_*}$ $<$ 3.6 \kms.

\subsection{Spectral energy distribution analysis}
\label{sec:SED_Analysis}

\begin{figure}
  \centering
  \includegraphics[angle=90, width=\columnwidth]{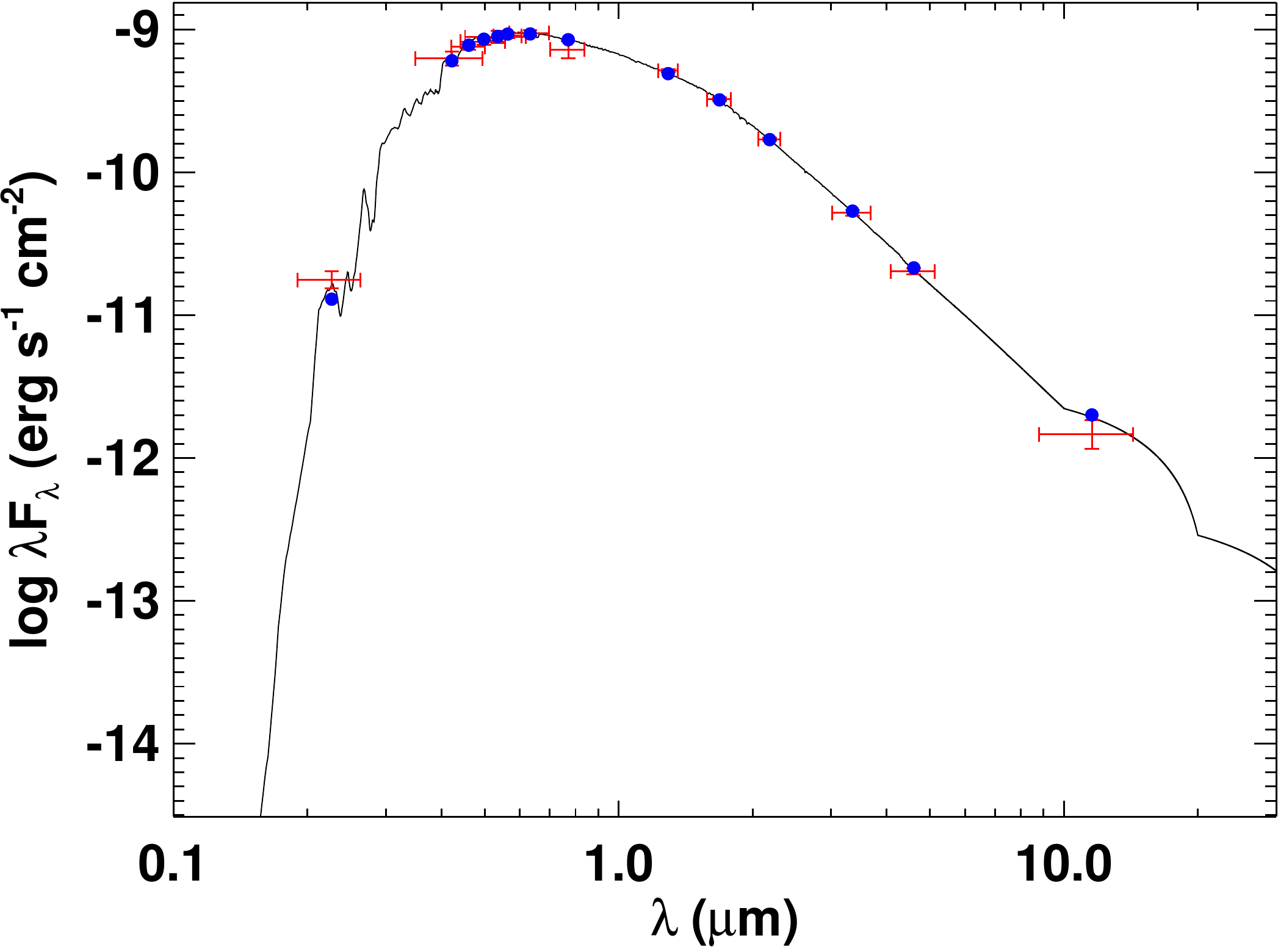}
  \caption{Measured and best fitting spectral energy distribution (SED) for KELT-10 from ultraviolet through mid-infrared. The red error bars indicate measurements of the flux of KELT-10 in ultraviolet, optical, near-infrared  and mid-infrared passbands as listed in \autoref{tbl:Host_Lit_Props}. The vertical bars are the 1$\sigma$ photometric uncertainties, whereas the horizontal error bars are the effective widths of the passbands. The solid curve is the best fitting theoretical SED from the Kurucz models \citep{Castelli:2004}, assuming stellar parameters \teff, $\log{g_*}$ and \feh\space fixed at the adopted values in \autoref{tbl:KELT-10b_global_fit_properties_YY_Isochrones}, with A$_V$ and $d$ allowed to vary. The blue dots are the predicted passband integrated fluxes of the best fitting theoretical SED corresponding to our observed photometric bands. (A colour version of this figure is available in the online journal.)}
  \label{fig:SED_figure}
\end{figure}
We construct an empirical spectral energy distribution (SED) of KELT-10 using all available broadband photometry in the literature, shown in \autoref{fig:SED_figure}. We use the near-UV flux from GALEX \citep{Martin:2005}, the $B_T$ and $V_T$ fluxes from the Tycho-2 catalogue, $B$, $V$, $g'$, $r'$, and $i'$ fluxes from the AAVSO APASS catalogue, NIR fluxes in the $J$, $H$, and $K_S$ bands from the 2MASS Point Source Catalogue \citep{Cutri:2003, Skrutskie:2006}, and near-and mid-infrared fluxes in the WISE passbands \citep{Wright:2010}.


We fit these fluxes using the Kurucz atmosphere models \citep{Castelli:2004} by fixing the values of \teff, $\log{g_*}$ and \feh\space inferred from the global fit to the light curve and RV data as described in \autoref{sec:Global_Modeling} and listed in \autoref{tbl:CORALIE_RV}, and then finding the values of the visual extinction $A_V$ and distance $d$ that minimize $\chi^2$, with a maximum permitted $A_V$ of 0.17 based on the full line-of-sight extinction from the dust maps of \citet{Schlegel:1998}.

We find A$_V = 0.17^{+0.00}_{-0.16}$ and $d$ = 183$\pm14$ pc with the best fit model having a reduced $\chi^2 = 1.34$. This implies a very good quality of fit and further corroborates the final derived stellar parameters for the KELT-10 host star. We note that the quoted statistical uncertainties on A$_V$ and $d$ are likely to be underestimated because we have not accounted for the uncertainties in \teff, $\log{g_*}$ and \feh\space used to derive the model SED. Furthermore, it is likely that alternate model atmospheres would predict somewhat different SEDs and thus values of extinction and distance. 

Nonetheless, the best fitting SED model places the system at a vertical height of only $\sim$60 pc, well within the local dust disk scale-height of 125 pc \citep{Marshall:2006}, and the system is seen through just over half of the local column of extinction. 

\subsection{UVW space motion}
\label{sec:UVW_Space_Motion}

Determining of the three-dimensional velocity of an object as it passes through the Galaxy can help to assess its age and membership in stellar groups or associations, especially if it belongs to an identifiable kinematic stream. Following the procedures detailed in \citet{Johnson:1987}, we calculate the UVW space motion of the KELT-10 system and examine its potential membership in known stellar groups or kinematic streams in the solar neighbourhood.

Employing the KELT-10 astrometric and~2-dimensional kinematic data detailed in \autoref{tbl:Host_Lit_Props}, as well as its centre-of-mass $\gamma$ velocity listed in \autoref{tbl:KELT-10b_global_fit_properties_YY_Isochrones}, only a precise and accurate distance is further required in order to calculate its UVW space motion. Because the KELT-10 system is too faint to have been observed by the \textit{Hipparcos} mission, its trigonometric parallax is not known and we use the distance $d$ = 183$\pm14$ derived from the SED analysis in \autoref{sec:SED_Analysis}.

For the host star of KELT-10b, we calculate its space motions to be U~=~34.6~$\pm$~0.5~\kms, V~=~-2.3~$\pm$~1.3~\kms\space and W~=~-8.4~$\pm$~0.9~\kms\space (with positive U pointing toward the Galactic centre). In the calculation of the UVW space motion, we have corrected for the peculiar velocity of the Sun with respect to the local standard of rest (LSR) using the values of U~=~8.5~\kms, V~=~13.38~\kms\space and W~=~6.49~\kms\space from \cite{Coskunoglu:2011}.

In U-V space, KELT-10 is not consistent with the kinematics of young disk objects, nor with the Pleiades, Hyades, and Ursa Major groups \citep{Montes:2001, Seifahrt:2010}. In fact, the same lack of kinematic membership is also true of the AB Dor, Argus, Beta Pic, Carina, Columba, Tuc-Hor, TW Hya groups \citep{Bowler:2013}. In V-W space however, KELT-10 is border-line consistent with the Beta Pic group \citep{Bowler:2013} and the local young disk population \citep{Seifahrt:2010}. However, our age estimate for KELT-10 (4.5 $\pm$ 0.7 Gyr) makes it unlikely to be a member of the young disk population.


We therefore posit that KELT-10's UVW space motion does not place it among any of the local, young kinematic moving groups of stars, and its space motion appears unremarkable. The UVW space motion is consistent with a thin disk source at $\sim$99\% probability, according to the criteria set out in \cite{Bensby:2003}.

\subsection{Stellar models and age}
\label{sec:Stellar_Age}

\begin{figure}
  \includegraphics[angle=90, width=\columnwidth]{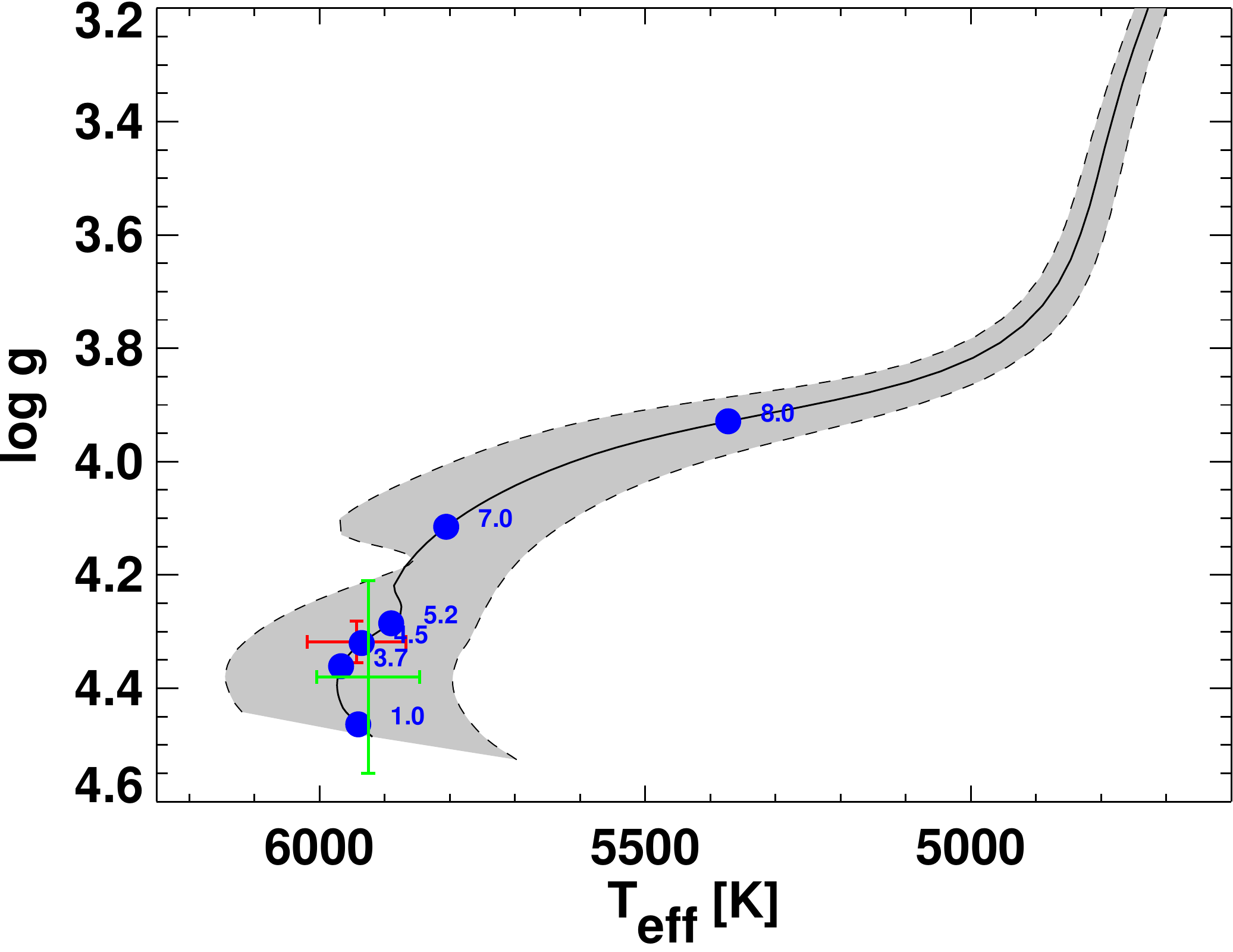}
  \caption{Theoretical H-R diagrams based on Yonsei-Yale stellar evolution models \citep{Demarque:2004}. The grey swaths represent the evolutionary track for the best-fit values of the mass and metallicity of the host star from the global fits corresponding to \autoref{tbl:KELT-10b_global_fit_properties_YY_Isochrones}. The tracks for the extreme range of 1$\sigma$ uncertainties on $M_{*}$ and \feh\space are shown as dashed lines bracketing each grey swath. The red cross is the position of KELT-10 using the values obtained from the best fitting global model and the blue points label various ages along the evolutionary track. The green cross is the position of KELT-10 using only the spectroscopically determined $\log{g_*}$. (A colour version of this figure is available in the online journal.)}
  \label{fig:hrd}
\end{figure}

To better place the KELT-10 system in context, we show in \autoref{fig:hrd} the H-R diagram for KELT-10 in the \teff\space vs.\ $\log{g_*}$ plane. We use the Yonsei-Yale stellar evolution model  track \citep{Demarque:2004} for a star with the mass and metallicity inferred from the global fit, where the shaded region represents the mass and \feh\space fit uncertainties. The model isochrone ages are indicated as blue points, and the final best global fit \teff\space and $\log{g_*}$ values are represented by the red error bars. For comparison, the \teff\space and $\log{g_*}$ values determined from spectroscopy alone are represented by the green error bars. KELT-10 is evidently a G0V star \citep{Pickles:2010} with an apparent age of $\sim$ 4.5$\pm$0.7 Gyr.


\section{Planetary properties}
\label{sec:Planet_properties}

\subsection{Global modelling with \textsc{EXOFAST}}
\label{sec:Global_Modeling}

Using a modified version of the IDL exoplanet fitting tool, \textsc{EXOFAST} \citep{Eastman:2013}, we perform a global fit of both the photometric and spectroscopic data described in \autoref{sec:KS_Photometry}, \autoref{sec:Follow-up_Photometry} and \autoref{sec:Spec_Follow-up}. This process is described further in \citet{Siverd:2012}. Using the Yonsei-Yale stellar evolution models \citep{Demarque:2004} or the empirical Torres relations \citep{Torres:2010} to constrain M$_{\star}$ and R$_{\star}$, \textsc{EXOFAST} performs a simultaneous Markov Chain Monte Carlo (MCMC) analysis of the spectroscopic data from CORALIE and follow-up photometric observations.

The raw light curve data and the detrending parameters (determined in \autoref{sec:Follow-up_Photometry}) were provided as inputs to \textsc{EXOFAST}, which performed the final detrending as part of the global fit.

We performed two global fits using the Yonsei-Yale stellar evolutionary models; one in which we allowed for an eccentric orbit (we did not fit for a long term slope in the radial velocity data since none was detected in a preliminary radial velocity analysis) and the second with a fixed circular orbit. The global fit included priors on P and T$_{c}$ from the KELT-South photometry, \teff, \feh, and $v\sin{I_*}$ from the SME analysis of the CORALIE spectrum. We imposed a starting point for $\log{g_*}$ using the spectroscopically determined value, but it was left as a free parameter during the global fitting procedure.

The median parameter values and inferred uncertainties from the global fits using the Yonsei-Yale stellar evolutionary models for the eccentric and circular orbit are compared in \autoref{tbl:KELT-10b_global_fit_properties_YY_Isochrones}. Similarly we performed two global fits using the Torres relations with the same priors as before. The median parameter values and inferred uncertainties are listed in \autoref{tbl:KELT-10b_global_fit_properties_Torres}. The four separate global fit parameters are consistent with each other to 1$\sigma$.

\begin{table*}
 \centering
  \caption{Median values and 68\% confidence interval for the physical and orbital parameters of the KELT-10 system using the YY models.}
  \label{tbl:KELT-10b_global_fit_properties_YY_Isochrones}
  \begin{tabular}{lccc}
  \hline
  \hline
   Parameter & Units & \textbf{Adopted Value} & Value \\
   & & \textbf{(YY circular)} & (YY eccentric)\\
 \hline
 Stellar Parameters: & & &\\
                          ~~~$M_{*}$\dotfill &Mass ($M_{\sun}$)\dotfill & $1.112_{-0.061}^{+0.055}$ 		& $1.112_{-0.059}^{+0.057}$ 		\\
                        ~~~$R_{*}$\dotfill &Radius ($R_{\sun}$)\dotfill & $1.209_{-0.035}^{+0.047}$ 		& $1.204_{-0.045}^{+0.054}$ 		\\
                    ~~~$L_{*}$\dotfill &Luminosity ($L_{\sun}$)\dotfill & $1.65_{-0.14}^{+0.17}$ 		& $1.63_{-0.16}^{+0.19}$ 		\\
                             ~~~$\rho_*$\dotfill &Density (cgs)\dotfill & $0.889_{-0.088}^{+0.062}$ 		& $0.901_{-0.10}^{+0.089}$ 		\\
                  ~~~$\log{g_*}$\dotfill &Surface gravity (cgs)\dotfill & $4.319_{-0.030}^{+0.020}$ 		& $4.323_{-0.035}^{+0.029}$ 		\\
                    ~~~\teff\dotfill &Effective temperature (K)\dotfill & $5948\pm74$ 				& $5948_{-76}^{+74}$ 			\\
                                   ~~~\feh\dotfill &Metallicity\dotfill & $0.09_{-0.10}^{+0.11}$ 		& $0.10_{-0.10}^{+0.11}$ 		\\
 \hline
Planetary Parameters: & & &\\
                                   ~~~$e$\dotfill &Eccentricity\dotfill & --- 					& $0.021_{-0.015}^{+0.023}$ 		\\
        ~~~$\omega_*$\dotfill &Argument of periastron (degrees)\dotfill & --- 					& $-144_{-90}^{+69}$ 			\\
                         ~~~$P$\dotfill &Period (days)$^{\dag}$\dotfill & $4.166285\pm0.000057$ 		& $4.166284_{-0.000057}^{+0.000058}$ 	\\
                           ~~~$a$\dotfill &Semi-major axis (AU)\dotfill & $0.05250_{-0.00097}^{+0.00086}$ 	& $0.05250_{-0.00095}^{+0.00088}$ 	\\
                                 ~~~$M_{P}$\dotfill &Mass (\MJ)\dotfill & $0.679_{-0.038}^{+0.039}$ 		& $0.679_{-0.034}^{+0.035}$ 		\\
                               ~~~$R_{P}$\dotfill &Radius (\RJ)\dotfill & $1.399_{-0.049}^{+0.069}$ 		& $1.392_{-0.059}^{+0.075}$ 		\\
                           ~~~$\rho_{P}$\dotfill &Density (cgs)\dotfill & $0.308_{-0.040}^{+0.033}$ 		& $0.312_{-0.045}^{+0.041}$ 		\\
                      ~~~$\log{g_{P}}$\dotfill &Surface gravity\dotfill & $2.933_{-0.041}^{+0.032}$ 		& $2.938_{-0.045}^{+0.037}$ 		\\
               ~~~$T_{eq}$\dotfill &Equilibrium temperature (K)\dotfill & $1377_{-23}^{+28}$ 			& $1373_{-29}^{+32}$ 			\\
                           ~~~$\Theta$\dotfill &Safronov number\dotfill & $0.0457\pm0.0027$ 			& $0.0459_{-0.0028}^{+0.0027}$		\\
                   ~~~$\fave$\dotfill &Incident flux (\fluxcgs)\dotfill & $0.817_{-0.054}^{+0.068}$ 		& $0.808_{-0.067}^{+0.077}$ 		\\
\hline
RV Parameters: & & &\\
~~~$T_C$\dotfill &Time of inferior conjunction (\bjdtdb)$^{\dag}$\dotfill & $2457175.0439_{-0.0054}^{+0.0053}$ 	& $2457175.0442_{-0.0053}^{+0.0054}$ 	\\
               ~~~$T_{P}$\dotfill &Time of periastron (\bjdtdb)\dotfill & --- 					& $2457176.45_{-1.0}^{+0.82}$ 		\\
                        ~~~$K$\dotfill &RV semi-amplitude (\ms)\dotfill & $80.0_{-3.5}^{+3.4}$ 			& $79.9\pm2.9$ 				\\
                    ~~~$M_P\sin{i}$\dotfill &Minimum mass (\MJ)\dotfill & $0.679_{-0.038}^{+0.039}$ 		& $0.678_{-0.034}^{+0.035}$ 		\\
                           ~~~$M_{P}/M_{*}$\dotfill &Mass ratio\dotfill & $0.000584\pm0.000027$ 		& $0.000583_{-0.000023}^{+0.000024}$ 	\\
                       ~~~$u$\dotfill &RM linear limb darkening\dotfill & $0.6387_{-0.0100}^{+0.010}$ 		& $0.6391_{-0.0100}^{+0.010}$ 		\\
                           ~~~$\gamma_{CORALIE}$\dotfill &(\ms)\dotfill & $31900\pm100$ 			& $31906\pm87$ 				\\
                     ~~~$\dot{\gamma}$\dotfill &RV slope (\msd)\dotfill & $-0.06_{-0.45}^{+0.44}$ 		& $-0.05\pm0.38$ 			\\
                                         ~~~$\ecosw$\dotfill & \dotfill & --- 					& $-0.008_{-0.020}^{+0.012}$ 		\\
                                         ~~~$\esinw$\dotfill & \dotfill & --- 					& $-0.003_{-0.026}^{+0.016}$ 		\\
\hline
Linear Ephemeris from & & & \\
Follow-up Transits: & & & \\
~~~$P_{Trans}$\dotfill &Period (days)\dotfill & $4.1662739\pm0.0000063$ & --- \\
~~~$T_{0}$\dotfill &Linear ephemeris from transits (\bjdtdb)\dotfill & $2457066.72045\pm0.00027$ & ---  \\
 \hline
 \hline
\end{tabular}
\begin{flushleft}
  \footnotesize \textbf{\textsc{NOTES}} \\
  \footnotesize $^{\dag}$ These values are less precise as they do not make use of the follow-up transit data during calculation.
  \end{flushleft} 
\end{table*}

\begin{table*}
 \centering
  \contcaption{Median values and 68\% confidence interval for the physical and orbital parameters of the KELT-10 system using the YY models.}
  \label{tbl:KELT-10b_global_fit_properties_YY_Isochrones_continued}
  \begin{tabular}{lccc}
  \hline
  \hline
   Parameter & Units & \textbf{Adopted Value} & Value \\
   & & \textbf{(YY circular)} & (YY eccentric)\\
 \hline
Primary Transit Parameters: & & & \\
~~~$R_{P}/R_{*}$\dotfill &Radius of the planet in stellar radii\dotfill & $0.1190_{-0.0012}^{+0.0014}$ 		& $0.1190_{-0.0012}^{+0.0014}$ 		\\
           ~~~$a/R_*$\dotfill &Semi-major axis in stellar radii\dotfill & $9.34_{-0.32}^{+0.21}$ 		& $9.39_{-0.37}^{+0.30}$ 		\\
                          ~~~$i$\dotfill &Inclination (degrees)\dotfill & $88.61_{-0.74}^{+0.86}$ 		& $88.63_{-0.76}^{+0.85}$ 		\\
                               ~~~$b$\dotfill &Impact parameter\dotfill & $0.23_{-0.14}^{+0.11}$ 		& $0.23_{-0.14}^{+0.11}$ 		\\
                             ~~~$\delta$\dotfill &Transit depth\dotfill & $0.01416_{-0.00029}^{+0.00034}$ 	& $0.01415_{-0.00029}^{+0.00034}$ 	\\
                    ~~~$T_{FWHM}$\dotfill &FWHM duration (days)\dotfill & $0.13848\pm0.00065$ 			& $0.13844\pm0.00066$ 			\\
              ~~~$\tau$\dotfill &Ingress/egress duration (days)\dotfill & $0.01743_{-0.00087}^{+0.0014}$ 	& $0.01742_{-0.00088}^{+0.0015}$ 	\\
                     ~~~$T_{14}$\dotfill &Total duration (days)\dotfill & $0.1560_{-0.0012}^{+0.0016}$ 		& $0.1559_{-0.0012}^{+0.0017}$ 		\\
   ~~~$P_{T}$\dotfill &A priori non-grazing transit probability\dotfill & $0.0943_{-0.0020}^{+0.0032}$ 		& $0.0936_{-0.0047}^{+0.0048}$ 		\\
             ~~~$P_{T,G}$\dotfill &A priori transit probability\dotfill & $0.1197_{-0.0027}^{+0.0043}$ 		& $0.1189_{-0.0060}^{+0.0064}$ 		\\
             ~~~$u_{1B}$\dotfill &Linear Limb-darkening\dotfill 	& $0.618_{-0.022}^{+0.023}$ 		& $0.618_{-0.022}^{+0.023}$ 		\\
                  ~~~$u_{2B}$\dotfill &Quadratic Limb-darkening\dotfill & $0.184_{-0.017}^{+0.016}$ 		& $0.183_{-0.017}^{+0.016}$ 		\\
                     ~~~$u_{1R}$\dotfill &Linear Limb-darkening\dotfill & $0.340\pm0.014$ 			& $0.341_{-0.014}^{+0.015}$ 		\\
                  ~~~$u_{2R}$\dotfill &Quadratic Limb-darkening\dotfill & $0.2958_{-0.0069}^{+0.0062}$ 		& $0.2956_{-0.0071}^{+0.0062}$ 		\\
                ~~~$u_{1Sloang}$\dotfill &Linear Limb-darkening\dotfill & $0.539_{-0.020}^{+0.021}$ 		& $0.540_{-0.020}^{+0.021}$ 		\\
             ~~~$u_{2Sloang}$\dotfill &Quadratic Limb-darkening\dotfill & $0.231_{-0.014}^{+0.013}$ 		& $0.231_{-0.014}^{+0.013}$ 		\\
                ~~~$u_{1Sloani}$\dotfill &Linear Limb-darkening\dotfill & $0.282\pm0.012$ 			& $0.283\pm0.012$ 			\\
             ~~~$u_{2Sloani}$\dotfill &Quadratic Limb-darkening\dotfill & $0.2916_{-0.0054}^{+0.0049}$ 		& $0.2915_{-0.0055}^{+0.0049}$ 		\\
                ~~~$u_{1Sloanz}$\dotfill &Linear Limb-darkening\dotfill & $0.2259_{-0.0097}^{+0.010}$ 		& $0.2262_{-0.0097}^{+0.010}$ 		\\
             ~~~$u_{2Sloanz}$\dotfill &Quadratic Limb-darkening\dotfill & $0.2854_{-0.0044}^{+0.0042}$ 		& $0.2853_{-0.0045}^{+0.0042}$ 		\\
 \hline
Secondary Eclipse Parameters: & & & \\
                  ~~~$T_{S}$\dotfill &Time of eclipse (\bjdtdb)\dotfill & $2457172.9607_{-0.0054}^{+0.0053}$ 	& $2457177.105_{-0.053}^{+0.032}$ 	\\
                           ~~~$b_{S}$\dotfill &Impact parameter\dotfill & --- 					& $0.22_{-0.14}^{+0.11}$      		\\
                  ~~~$T_{S,FWHM}$\dotfill &FWHM duration (days)\dotfill & --- 					& $0.1375_{-0.0065}^{+0.0042}$		\\
            ~~~$\tau_S$\dotfill &Ingress/egress duration (days)\dotfill & --- 					& $0.0172_{-0.0011}^{+0.0017}$		\\
                   ~~~$T_{S,14}$\dotfill &Total duration (days)\dotfill & --- 					& $0.1549_{-0.0074}^{+0.0052}$		\\
   ~~~$P_{S}$\dotfill &A priori non-grazing eclipse probability\dotfill & --- 					& $0.0944_{-0.0020}^{+0.0033}$		\\
             ~~~$P_{S,G}$\dotfill &A priori eclipse probability\dotfill & --- 					& $0.1198_{-0.0028}^{+0.0045}$		\\
 \hline
 \hline
\end{tabular}
\end{table*}

\begin{table*}
 \centering
  \caption{Median values and 68\% confidence interval for the physical and orbital parameters of the KELT-10 system using the Torres relations.}
  \label{tbl:KELT-10b_global_fit_properties_Torres}
  \begin{tabular}{lccc}
  \hline
  \hline
   Parameter & Units & Value & Value \\
   & & (Torres circular) & (Torres eccentric)\\
 \hline
 Stellar Parameters: & & &\\
                          ~~~$M_{*}$\dotfill &Mass ($M_{\sun}$)\dotfill & $1.143_{-0.059}^{+0.060}$ 		& $1.140_{-0.059}^{+0.061}$ 		\\
                        ~~~$R_{*}$\dotfill &Radius ($R_{\sun}$)\dotfill & $1.221_{-0.035}^{+0.047}$ 		& $1.215_{-0.047}^{+0.055}$ 		\\
                    ~~~$L_{*}$\dotfill &Luminosity ($L_{\sun}$)\dotfill & $1.69_{-0.14}^{+0.17}$ 		& $1.66_{-0.16}^{+0.19}$ 		\\
                             ~~~$\rho_*$\dotfill &Density (cgs)\dotfill & $0.889_{-0.084}^{+0.060}$ 		& $0.900_{-0.098}^{+0.086}$ 		\\
                  ~~~$\log{g_*}$\dotfill &Surface gravity (cgs)\dotfill & $4.323_{-0.028}^{+0.019}$ 		& $4.326_{-0.033}^{+0.027}$ 		\\
                    ~~~\teff\dotfill &Effective temperature (K)\dotfill & $5951_{-76}^{+75}$			& $5950_{-75}^{+76}$ 			\\
                                   ~~~\feh\dotfill &Metallicity\dotfill & $0.09\pm0.11$ 			& $0.09\pm0.11$ 			\\
 \hline
Planetary Parameters: & & &\\
                                   ~~~$e$\dotfill &Eccentricity\dotfill & ---					& $0.021_{-0.015}^{+0.023}$ 		\\
        ~~~$\omega_*$\dotfill &Argument of periastron (degrees)\dotfill & ---		 			& $-147_{-91}^{+72}$ 			\\
                         ~~~$P$\dotfill &Period (days)$^{\dag}$\dotfill & $4.166286\pm0.000057$ 		& $4.166283\pm0.000057$ 		\\
                           ~~~$a$\dotfill &Semi-major axis (AU)\dotfill & $0.05298_{-0.00093}^{+0.00091}$	& $0.05293_{-0.00092}^{+0.00093}$ 	\\
                                 ~~~$M_{P}$\dotfill &Mass (\MJ)\dotfill & $0.692_{-0.038}^{+0.039}$ 		& $0.689_{-0.034}^{+0.036}$ 		\\
                               ~~~$R_{P}$\dotfill &Radius (\RJ)\dotfill & $1.413_{-0.049}^{+0.068}$ 		& $1.405_{-0.061}^{+0.077}$ 		\\
                           ~~~$\rho_{P}$\dotfill &Density (cgs)\dotfill & $0.304_{-0.038}^{+0.032}$		& $0.308_{-0.043}^{+0.039}$ 		\\
                      ~~~$\log{g_{P}}$\dotfill &Surface gravity\dotfill & $2.933_{-0.040}^{+0.032}$ 		& $2.936_{-0.043}^{+0.036}$ 		\\
               ~~~$T_{eq}$\dotfill &Equilibrium temperature (K)\dotfill & $1378_{-24}^{+28}$ 			& $1374_{-29}^{+32}$ 			\\
                           ~~~$\Theta$\dotfill &Safronov number\dotfill & $0.0452_{-0.0027}^{+0.0026}$		& $0.0454_{-0.0028}^{+0.0027}$ 		\\
                   ~~~$\fave$\dotfill &Incident flux (\fluxcgs)\dotfill & $0.818_{-0.055}^{+0.068}$ 		& $0.810_{-0.066}^{+0.077}$ 		\\
\hline
RV Parameters: & & &\\
~~~$T_C$\dotfill &Time of inferior conjunction (\bjdtdb)$^{\dag}$\dotfill & $2457175.0438\pm0.0053$ 		& $2457175.0444\pm0.0053$ 		\\
               ~~~$T_{P}$\dotfill &Time of periastron (\bjdtdb)\dotfill & ---					& $2457176.41_{-1.0}^{+0.85}$ 		\\
                        ~~~$K$\dotfill &RV semi-amplitude (\ms)\dotfill & $80.0_{-3.4}^{+3.5}$			& $79.8\pm2.9$ 				\\
                    ~~~$M_P\sin{i}$\dotfill &Minimum mass (\MJ)\dotfill & $0.692_{-0.038}^{+0.039}$ 		& $0.689_{-0.034}^{+0.036}$ 		\\
                           ~~~$M_{P}/M_{*}$\dotfill &Mass ratio\dotfill & $0.000579_{-0.000026}^{+0.000027}$	& $0.000577\pm0.000023$ 		\\
                       ~~~$u$\dotfill &RM linear limb darkening\dotfill & $0.6382_{-0.0099}^{+0.010}$ 		& $0.6381_{-0.0100}^{+0.010}$ 		\\
                           ~~~$\gamma_{CORALIE}$\dotfill &(\ms)\dotfill & $31900\pm100$				& $31907_{-88}^{+87}$ 			\\
                     ~~~$\dot{\gamma}$\dotfill &RV slope (\msd)\dotfill & $-0.06\pm0.45$ 			& $-0.05\pm0.38$ 			\\
                                         ~~~$\ecosw$\dotfill & \dotfill & ---					& $-0.008_{-0.020}^{+0.012}$ 		\\
                                         ~~~$\esinw$\dotfill & \dotfill & ---					& $-0.003_{-0.025}^{+0.016}$ 		\\
 \hline
 \hline
\end{tabular}
\begin{flushleft}
  \footnotesize \textbf{\textsc{NOTES}} \\
  \footnotesize $^{\dag}$ These values are less precise as they do not make use of the follow-up transit data during calculation.
  \end{flushleft} 
\end{table*}

\begin{table*}
 \centering
  \contcaption{Median values and 68\% confidence interval for the physical and orbital parameters of the KELT-10 system using the Torres relations.}
  \label{tbl:KELT-10b_global_fit_properties_Torres_continued}
  \begin{tabular}{lccc}
  \hline
  \hline
   Parameter & Units & Value & Value \\
   & & (Torres circular) & (Torres eccentric)\\
 \hline
Primary Transit Parameters: & & & \\
~~~$R_{P}/R_{*}$\dotfill &Radius of the planet in stellar radii\dotfill &  $0.1191_{-0.0012}^{+0.0014}$		& $0.1190_{-0.0012}^{+0.0014}$ 		\\
           ~~~$a/R_*$\dotfill &Semi-major axis in stellar radii\dotfill &  $9.34_{-0.30}^{+0.20}$		& $9.38_{-0.35}^{+0.29}$ 		\\
                          ~~~$i$\dotfill &Inclination (degrees)\dotfill &  $88.63_{-0.72}^{+0.85}$		& $88.67_{-0.75}^{+0.84}$ 		\\
                               ~~~$b$\dotfill &Impact parameter\dotfill &  $0.22_{-0.14}^{+0.11}$		& $0.22_{-0.14}^{+0.11}$ 		\\
                             ~~~$\delta$\dotfill &Transit depth\dotfill &  $0.01418_{-0.00028}^{+0.00033}$	& $0.01417_{-0.00028}^{+0.00033}$ 	\\
                    ~~~$T_{FWHM}$\dotfill &FWHM duration (days)\dotfill &  $0.13859_{-0.00063}^{+0.00064}$	& $0.13859_{-0.00063}^{+0.00064}$ 	\\
              ~~~$\tau$\dotfill &Ingress/egress duration (days)\dotfill &  $0.01743_{-0.00085}^{+0.0014}$	& $0.01739_{-0.00083}^{+0.0014}$ 	\\
                     ~~~$T_{14}$\dotfill &Total duration (days)\dotfill &  $0.1561_{-0.0012}^{+0.0015}$		& $0.1561_{-0.0012}^{+0.0016}$ 		\\
   ~~~$P_{T}$\dotfill &A priori non-grazing transit probability\dotfill &  $0.0943_{-0.0019}^{+0.0030}$		& $0.0937_{-0.0045}^{+0.0047}$ 		\\
             ~~~$P_{T,G}$\dotfill &A priori transit probability\dotfill &  $0.1198_{-0.0026}^{+0.0042}$		& $0.1190_{-0.0058}^{+0.0062}$ 		\\
             ~~~$u_{1B}$\dotfill &Linear Limb-darkening\dotfill 	&  $0.617_{-0.022}^{+0.023}$		& $0.616_{-0.022}^{+0.023}$ 		\\
               ~~~$u_{2B}$\dotfill &Quadratic Limb-darkening\dotfill 	&  $0.185_{-0.017}^{+0.016}$		& $0.185_{-0.017}^{+0.016}$ 		\\
                     ~~~$u_{1R}$\dotfill &Linear Limb-darkening\dotfill &  $0.340_{-0.014}^{+0.015}$		& $0.340\pm0.014$ 			\\
                  ~~~$u_{2R}$\dotfill &Quadratic Limb-darkening\dotfill &  $0.2960_{-0.0071}^{+0.0062}$		& $0.2959_{-0.0069}^{+0.0062}$ 		\\
                ~~~$u_{1Sloang}$\dotfill &Linear Limb-darkening\dotfill &  $0.538_{-0.020}^{+0.021}$		& $0.538_{-0.020}^{+0.021}$		\\
             ~~~$u_{2Sloang}$\dotfill &Quadratic Limb-darkening\dotfill &  $0.232_{-0.014}^{+0.013}$		& $0.232_{-0.014}^{+0.013}$ 		\\
                ~~~$u_{1Sloani}$\dotfill &Linear Limb-darkening\dotfill &  $0.282\pm0.012$			& $0.282\pm0.012$ 			\\
             ~~~$u_{2Sloani}$\dotfill &Quadratic Limb-darkening\dotfill &  $0.2918_{-0.0055}^{+0.0049}$		& $0.2917_{-0.0054}^{+0.0049}$ 		\\
                ~~~$u_{1Sloanz}$\dotfill &Linear Limb-darkening\dotfill &  $0.2255_{-0.0097}^{+0.010}$		& $0.2255_{-0.0098}^{+0.010}$ 		\\
             ~~~$u_{2Sloanz}$\dotfill &Quadratic Limb-darkening\dotfill &  $0.2855_{-0.0045}^{+0.0042}$		& $0.2854_{-0.0045}^{+0.0042}$ 		\\
 \hline
Secondary Eclipse Parameters: & & & \\
                  ~~~$T_{S}$\dotfill &Time of eclipse (\bjdtdb)\dotfill &  $2457172.9607\pm0.0053$		& $2457177.106_{-0.053}^{+0.032}$ 	\\
                           ~~~$b_{S}$\dotfill &Impact parameter\dotfill & ---					& $0.22_{-0.13}^{+0.11}$ 		\\
                  ~~~$T_{S,FWHM}$\dotfill &FWHM duration (days)\dotfill & ---					& $0.1378_{-0.0063}^{+0.0043}$ 		\\
            ~~~$\tau_S$\dotfill &Ingress/egress duration (days)\dotfill & ---					& $0.0173_{-0.0011}^{+0.0016}$ 		\\
                   ~~~$T_{S,14}$\dotfill &Total duration (days)\dotfill & ---					& $0.1552_{-0.0072}^{+0.0052}$ 		\\
   ~~~$P_{S}$\dotfill &A priori non-grazing eclipse probability\dotfill & ---					& $0.0943_{-0.0019}^{+0.0032}$ 		\\
             ~~~$P_{S,G}$\dotfill &A priori eclipse probability\dotfill & ---					& $0.1198_{-0.0026}^{+0.0043}$ 		\\
 \hline
 \hline
\end{tabular}
\end{table*}

\subsection{Transit timing variation analysis}
\label{sec:TTV}

\begin{table*}
 \centering
 \caption{Transit times for KELT-10b.}
 \label{tbl:transitimes}
 \begin{tabular}{llllll}
    \hline
    \hline
    Epoch & $T_{C}$ 	& $\sigma_{T_{C}}$ 	& O-C &  O-C 			& Telescope \\
	  & (\bjdtdb) 	& (s) 			& (s) & ($\sigma_{T_{C}}$) 	& \\
    \hline
 -54 & 2456841.741248 &  37 &  -35.71 & -0.96 & LSC \\
 -52 & 2456850.074699 &  69 &   42.32 &  0.61 & PEST \\
 -46 & 2456875.074541 & 105 &  232.27 &  2.19 & PEST \\
  26 & 2457175.045092 &  82 &  131.13 &  1.59 & COJ \\
  26 & 2457175.041903 & 104 & -144.39 & -1.39 & MTKENT \\
  27 & 2457179.207605 & 101 & -193.81 & -1.91 & ICO \\
  27 & 2457179.208733 &  56 &  -96.35 & -1.72 & MTKENT \\
  30 & 2457191.709930 &  52 &  108.87 &  2.09 & PROMPT5 \\
    \hline
    \hline
 \end{tabular}
\end{table*} 

\begin{figure}
  \includegraphics[width=\columnwidth]{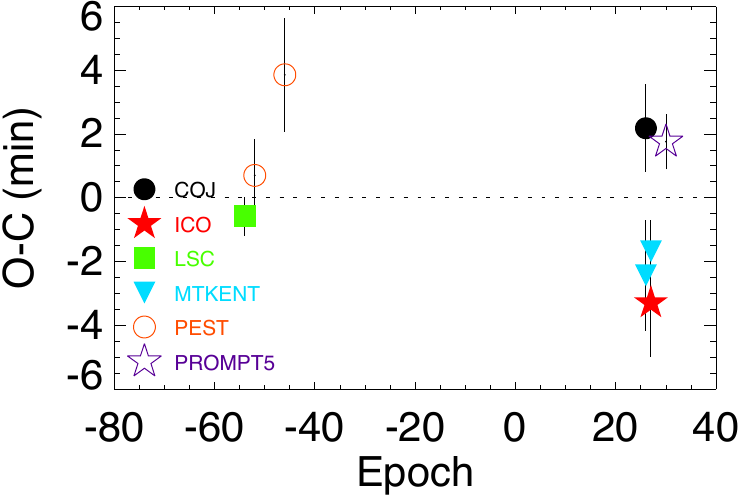}
  \caption{Residuals of the transit times from the best-fit ephemeris. The transit times are given in \autoref{tbl:transitimes}. The observatory/telescope abbreviations are the same as in \autoref{fig:All_Lightcurve}. (A colour version of this figure is available in the online journal.)}
  \label{fig:ttv}
\end{figure}
We searched for possible transit timing variations (TTVs) in the system by allowing the transit times for each of the follow-up light curves to vary. We were careful to ensure that all quoted times had been properly reported in \bjdtdb\space \citep[e.g.,][]{Eastman:2010}. The ephemeris quoted in \autoref{tbl:KELT-10b_global_fit_properties_YY_Isochrones} and \autoref{tbl:KELT-10b_global_fit_properties_Torres} is only constrained by the RV data and a prior imposed from the KELT-South discovery data. Using the follow-up transit light curves to constrain the ephemeris in the global fit would require us to adopt a specific model for the parameters of any perturbing body, which would be poorly constrained and highly degenerate, or it would require us to assume a linear ephemeris, thereby eliminating the possibility of any TTV signal altogether.

Subsequent to the global fit, we then derived a separate ephemeris from only the transit timing data by fitting a straight line to all inferred transit centre times from the global fit. These data are presented in \autoref{tbl:transitimes} and \autoref{fig:ttv}. We find $T_{0}$~=~2457066.72045$\pm$0.00027\space\bjdtdb\space and P~=~4.1662739$\pm$0.0000063 days, with a $\chi^2$ of 21.5 and 6 degrees of freedom. While chi$^2$/dof formally indicates a poor fit to the TTV data, we find that this is often the case in ground-based TTV studies, likely due to systematics in the transit data. Given the likely difficulty with properly removing systematics in partial transit data, we are unwilling to claim convincing evidence for TTVs. Although it is known that Hot Jupiters rarely have nearby companions in orbits that can induce TTVs \citep{Steffen:2007}, the case of WASP-47 \citep{Becker:2015} demonstrates that, at least some rare cases, such companions can and do exist. Therefore, we are unwilling to conclusively rule out a true astrophysical cause for the TTVs that we detect, and suggest that further observations of KELT-10b's transit times are required to definitively conclude (or refute) that systematics are the cause of the observed poor fit we find to a linear emphemeris.

\section{False positive analysis}
\label{sec:FP_Analysis}
We are confident that KELT-10b is a true planetary system but many astrophysical and non-astrophysical scenarios can also resemble a planetary signal. Here we discuss some false positive scenarios and how we exclude them in this instance.

The KELT-South telescope has a very large plate scale, 23\appx, and a large PSF and as a result, many of our candidates are blended with a nearby eclipsing binary. Using our follow-up network of telescopes, which have higher spatial resolution (PSF $\sim$ 1\arcsec), we confirm which star is displaying the transit signature. The follow-up observations of KELT-10 were in multiple bandpasses ($B$, $R$, $g^{\prime}$, $i^{\prime}$, $z^{\prime}$) and found no wavelength dependency in the transit depth, indicating the transit is due to a dark (or at least dim) object, i.e., a planet.

All spectroscopic observations of KELT-10 were carefully analysed to look for any signs of light from another source. The spectra show no evidence of multiple sets of absorption lines. Also, our bisector analysis (\autoref{fig:RV_BS}) shows no correlation between the measured radial velocities from CORALIE and the bisector spans. 

The AO observations detected a faint companion with a $\Delta K = 9$ to KELT-10. We are confident that the faint companion is neither capable of being the source nor able to affect the transit signal we see in KELT-10. The companion is 4000 times fainter than KELT-10 and even at a 100\% change in flux, the transit signal would be affected at a maximum level of 0.2 millimag, which is far below the precision we measure for the transit depth.

The $\log{g_*}$ derived from the global fit and the value derived from our SME analysis, are consistent within errors, having values of $4.319_{-0.030}^{+0.020}$ and $4.38\pm0.17$, respectively, and the global fit with all spectroscopic and photometric data is well modelled by that of a transiting planet around a single star. 

Therefore, we conclude that KELT-10b is a bona fide transiting planet.

\section{Discussion}
\label{sec:discussion_conclusion}

\subsection{Insolation evolution}
\label{sec:Insol_evol}

\begin{figure}
  \includegraphics[width=\columnwidth]{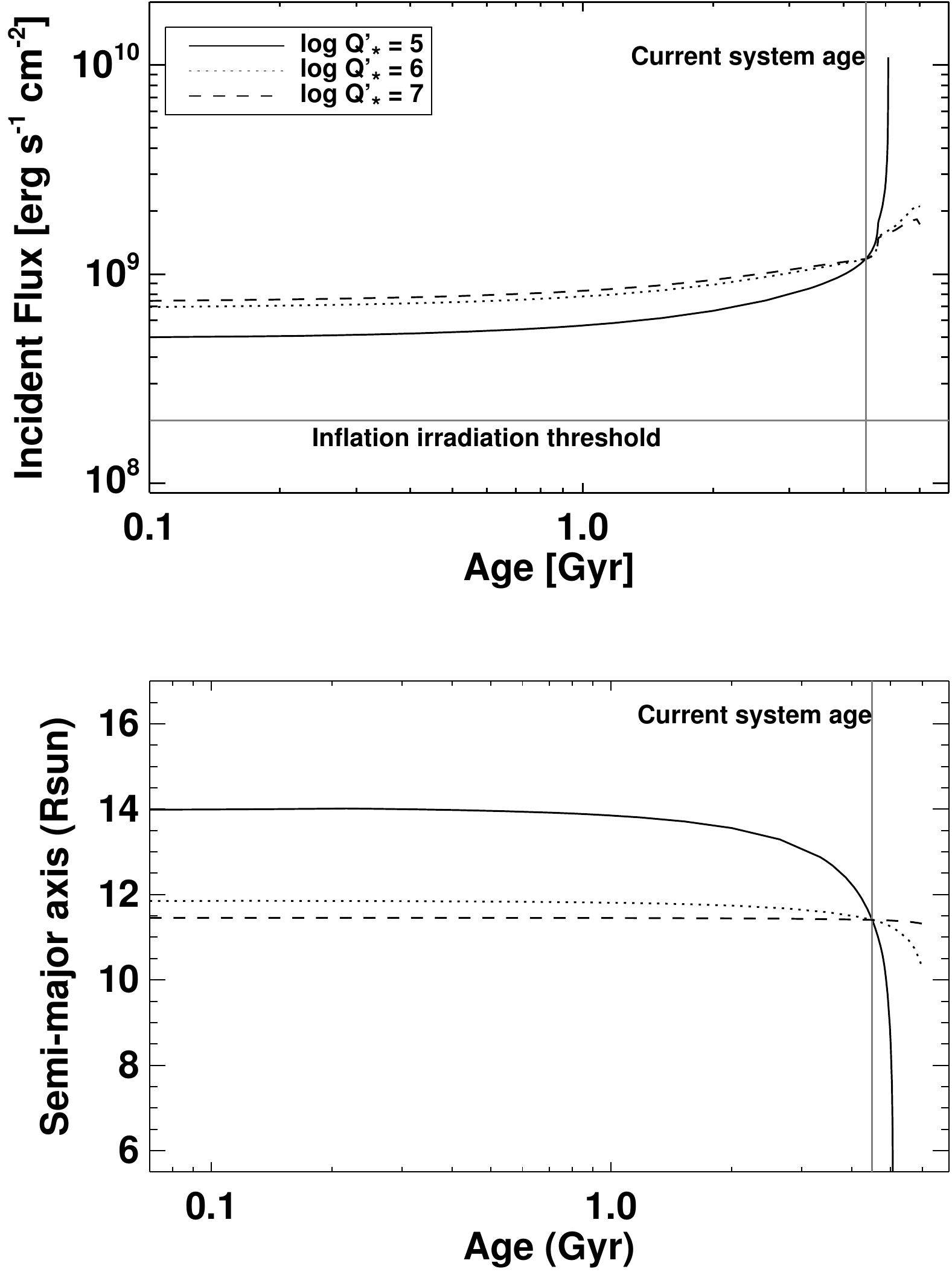}
  \caption{\textit{Top}: Irradiation history of KELT-10b, plotted for test values of $\log Q'_\star$ from 5 to 7. The insolation received by the planet is well above the empirical inflation irradiation threshold \citep{Demory:2011} for the entire main-sequence existence of the star and is clearly insensitive to the value of $Q'_\star$ in the range we analysed. \textit{Bottom}: Orbital semi-major axis history of KELT-10b. The planet's semi-major axis is rapidly decreasing as the star evolves off the main sequence. It appears unlikely that KELT-10b will survive past the star's current subgiant phase.}
  \label{fig:evol}
\end{figure}

As can be seen from the results of the global fit (\autoref{tbl:KELT-10b_global_fit_properties_YY_Isochrones}), KELT-10b is a highly inflated planet, joining the ranks of other hot Jupiters that manifest radii much larger than predicted by standard models for non-irradiated objects with Jovian masses. Several authors \citep[e.g.,][]{Demory:2011} have suggested an empirical insolation threshold ($\approx 2 \times 10^8$ erg s$^{-1}$ cm$^{-2}$) above which hot Jupiters exhibit increasing amounts of radius inflation. KELT-10b clearly lies above this threshold, with a current estimated insolation of $8.17_{-0.54}^{+0.68} \times 10^8$ erg s$^{-1}$ cm$^{-2}$ from the global fits, and therefore its currently large inflated radius is not surprising. At the same time, the KELT-10 host star is found to be at present in a state of evolution whereby its radius is expanding as the star crosses the Hertzsprung gap toward the red giant branch. This means that the star's surface is encroaching on the planet, which presumably is in turn driving up the planet's insolation and also the rate of any tidal interactions between the planet and the star. 

Therefore it is interesting to consider two questions. First, has KELT-10b's incident radiation from its host star been below the empirical radius inflation threshold in the past? If KELT-10b's insolation only recently exceeded the inflation threshold, the system could then serve as an empirical test bed for the different timescales predicted by different inflation mechanisms \citep[see, e.g.,][]{Assef:2009,Spiegel:2012}. Second, what is the expected fate of the KELT-10b planet given the increasingly strong tidal interactions it is experiencing with its encroaching host star? 

To investigate these questions, we follow \citet{Penev:2014} to simulate the reverse and forward evolution of the star-planet system, using the measured parameters listed in \autoref{tbl:KELT-10b_global_fit_properties_YY_Isochrones} as the present-day boundary conditions. This analysis is not intended to examine any type of planet-planet or planet-disk migration effects. Rather, it is a way to investigate (1) the change in insolation of the planet over time due to the changing luminosity of the star and changing star-planet separation, and (2) the change in the planet's orbital semi-major axis due to the changing tidal torque as the star-planet separation changes with the evolving stellar radius. We include the evolution of the star, assumed to follow the Yonsei-Yale stellar model with mass and metallicity as in \autoref{tbl:KELT-10b_global_fit_properties_YY_Isochrones}. For simplicity we assume that the stellar rotation is negligible and treat the star as a solid body. We also assume a circular orbit aligned with the stellar equator throughout the analysis. The results of our simulations are shown in \autoref{fig:evol}. We tested a range of values for the tidal quality factor of the star divided by the love number, $Q'_\star \equiv Q_\star / k_2$, from $\log Q'_\star = 5$ to $\log Q'_\star = 7$ (assuming a constant phase lag between the tidal bulge and the star-planet direction). We find that although for certain values of $Q'_\star$ the planet has moved substantially closer to its host during the past Gyr, in all cases the planet has always received more than enough flux from its host to keep the planet irradiated beyond the insolation threshold identified by \citet{Demory:2011}.

Interestingly, the current evolution of the star suggests a concomitant in-spiral of the planet over the next $\sim$1 Gyr, unless the stellar $Q'_\star$ is large. Therefore it is possible that this planet will not survive beyond the star's current subgiant phase. As additional systems like KELT-10b are discovered and their evolution investigated in detail, it will be interesting to examine the statistics of planet survival and to compare these to predictions such as those shown here in \autoref{fig:evol} to constrain mechanisms of planet-star interaction generally and the values of $Q'_\star$ specifically. Therefore it is possible that this planet will not survive beyond the star's current subgiant phase.

\subsection{Future characterization of the host star}
\label{sec:future_host_characterisation}
To better constrain the mass and radius of KELT-10, which will improve our knowledge of the planetary mass and radius, further stellar characterization is required. light curves with higher precision can be used to better constrain the stellar density. Higher S/N ratio spectra may help improve the spectroscopically determined parameters and allow for more constrained chemical abundance measurements. The GAIA mission \citep{deBruijne:2012} should provide a precise parallax measurement and more accurate values for the proper motion of KELT-10, which would enable better determinations of the UVW space motion of the system and allow a model-independent estimate of the properties of the host star and planet (i.e., that don't rely on YY-isochrones or the Torres relations). Long term follow-up imaging of KELT-10 would provide the necessary data to determine whether the faint companion seen in the AO images is bound to the system or not.

\subsection{Mass-Radius predictions}
\label{sec:MassRadius_predictions}
Several authors have identified empirical relations that attempt to predict the radius of a planet based on a combination of stellar and planetary properties. Empirical relations presented by \citet{Weiss:2013} use the incident flux and mass of the planet as dependent variables. Using equation 9 of \citet{Weiss:2013}, we determined the predicted radius of KELT-10b to be 1.22 \RJ when the value determined using the global fit is 1.399$_{-0.049}^{+0.069}$ \RJ\space ($\sim$15\% larger than predicted). If one takes into account the scatter of $\sim$0.1 \RJ\space in the \citet{Weiss:2013} relation, and the error on our radius measurement the predicted and measured values agree to within 1.1$\sigma$.

\subsection{Comparative planetology}
\label{sec:comparitive_planetology}

\begin{figure}
  \centering
  \includegraphics[width=\columnwidth]{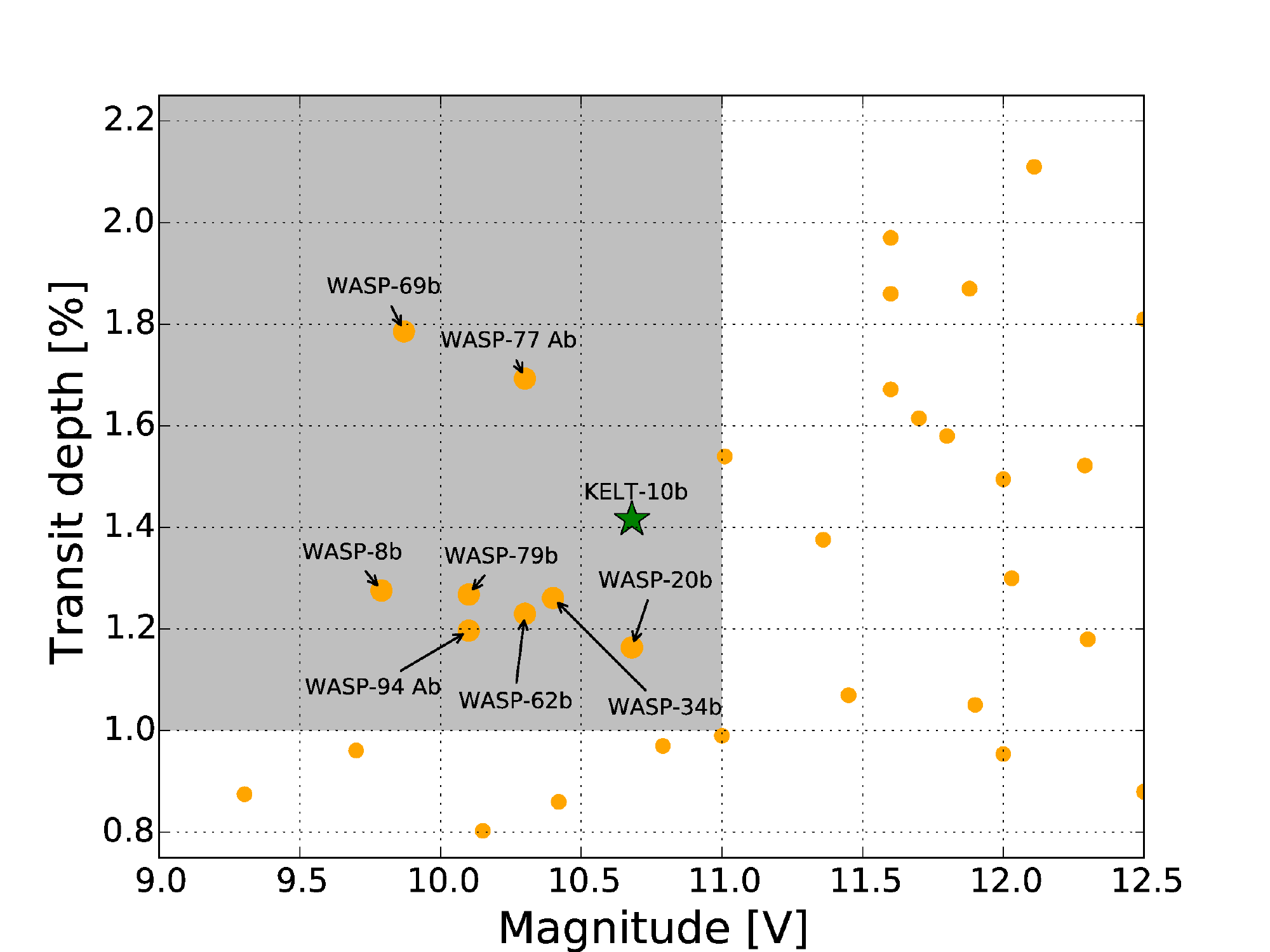}
  \caption{Transit depth as a function of the host star apparent V magnitude for southern transiting systems with relatively bright (V $<$ 12.5) host stars and relatively deep ($\delta >$ 0.75\%) transit depths. KELT-10b is shown as the green star. Systems in the top left (the grey shaded area) tend to be the most amenable to detailed spectroscopic and photometric studies. Other bright transiting exoplanet systems that fall into this category are also labelled (data obtained from the NASA Exoplanet Archive, retrieved on 2015 August 28). (A colour version of this figure is available in the online journal.)}
  \label{fig:trans_depth_vs_v_mag}
\end{figure}


KELT-10b is an inflated hot Jupiter, similar to a number of other such planets in the combination of mass and radius, especially HD209458b and HAT-P-9b. It is noteworthy for its combination of bright host star and transit depth (see \autoref{fig:trans_depth_vs_v_mag}), as well as mass and density of the planet. Since transiting planets with bright host stars are much more accessible for atmospheric studies, the V=10.7 brightness of KELT-10 places it among some of the best systems amenable for atmospheric characterization. With a transit depth of 1.4\%, KELT-10b has the third deepest transit among bright southern transiting planets with host star V $<$ 11, making it a good candidate for atmospheric analysis by large telescopes in the southern hemisphere. 

\section{Conclusion}
\label{sec:Conclusion}
We announce the discovery of a highly inflated hot sub-Jupiter, KELT-10b. It is the first transiting exoplanet discovered using the KELT-South survey. The planet was initially discovered in the KELT photometry, then confirmed via higher precision follow-up light curves and radial velocity measurements. KELT-10b, with \RP~=~$1.399_{-0.049}^{+0.069}$~\RJ\space and mass \MP~=~$0.679_{-0.038}^{+0.039}$~\MJ, joins a group of transiting exoplanets with highly inflated radii and masses below that of Jupiter. 

The planet transits a relatively bright star which is accessible to telescopes in the southern hemisphere and has the third deepest transit among southern transiting planets with host star V $<$ 11, making it a promising candidate for future atmospheric characterization studies.

\section*{Acknowledgements}
\label{sec:Acknowledgements}
KELT-South is hosted by the South African Astronomical Observatory and we are grateful for their ongoing support and assistance. K.P. acknowledges support from NASA grant NNX13AQ62G. Work by B.S.G. and D.J.S. was partially supported by NSF CAREER Grant AST-1056524. 

This research has made use of the Exoplanet Orbit Database and the Exoplanet Data Explorer at exoplanets.org \citep{Han:2014}. This work has made use of NASAs Astrophysics Data System, the Extrasolar Planet Encyclopaedia at exoplanet.eu \citep{Schneider:2011}, the SIMBAD database \citep{Wenger:2000} operated at CDS, Strasbourg, France, and the VizieR catalogue access tool, CDS, Strasbourg, France \citep{Ochsenbein:2000}. This paper makes use of data and services from NASA Exoplanet Archive \citep{Akeson:2013}, which is operated by the California Institute of Technology, under contract
with the National Aeronautics and Space Administration under the Exoplanet Exploration Program. This research has made use of TEPCat at http://www.astro.keele.ac.uk/jkt/tepcat/ \citep{Southworth:2011}.

This publication makes use of data products from the Wide-field Infrared Survey Explorer, which is a joint project of the University of California, Los Angeles, and the Jet Propulsion Laboratory/California Institute of Technology, funded by the National Aeronautics and Space Administration. This publication makes use of data products from the Two Micron All Sky Survey, which is a joint project of the University of Massachusetts and the Infrared Processing and Analysis Center/California Institute of Technology, funded by the National Aeronautics and Space Administration and the National Science Foundation.

This research was made possible through the use of the AAVSO Photometric All-Sky Survey (APASS), funded by the Robert Martin Ayers Sciences Fund. This paper uses observations obtained with facilities of the Las Cumbres Observatory Global Telescope.




\bibliographystyle{mnras}
\bibliography{KELT-10b}



%
%
%

\clearpage
\onecolumn

\noindent
Author affiliations:\\
\vspace{0.3cm}\\
\begin{footnotesize}
$^{1}$South African Astronomical Observatory, PO Box 9, Observatory 7935, South Africa\\
$^{2}$Department of Physics and Astronomy, Vanderbilt University, 6301 Stevenson Center, Nashville, TN 37235, USA\\
$^{3}$Department of Physics and Astronomy, University of Louisville, Louisville, KY 40292, USA\\
$^{4}$Las Cumbres Observatory Global Telescope Network, 6740 Cortona Drive, Suite 102, Santa Barbara, CA 93117, USA\\
$^{5}$Department of Physics, Lehigh University, Bethlehem, PA 18015, USA\\
$^{6}$NASA Ames Research Center, M/S 244-30, Moffett Field, CA 94035, USA\\
$^{7}$Bay Area Environmental Research Institute, 625 2nd St. Ste 209 Petaluma, CA 94952, USA\\
$^{8}$Department of Physics, Fisk University, 1000 17th Avenue North, Nashville, TN 37208, USA\\
$^{9}$Harvard-Smithsonian Center for Astrophysics, 60 Garden St, Cambridge, MA 02138, USA\\
$^{10}$Cerro Tololo Inter-American Observatory, Colina El Pino, S/N, La Serena, Chile.\\
$^{11}$Department of Astrophysical Sciences, Princeton University, Princeton, NJ 08544, USA\\
$^{12}$Research School of Astronomy and Astrophysics, Australian National University, Canberra, ACT 2611, Australia\\
$^{13}$Observatoire Astronomique de l'Universit\'{e} de Gen\`{e}ve, Chemin des Maillettes 51, 1290 Sauverny, Switzerland\\
$^{14}$The Australian National University, Canberra, Australia\\
$^{15}$Perth Exoplanet Survey Telescope, Perth, Australia\\
$^{16}$Adelaide, Australia\\
$^{17}$Department of Astronomy, California Institute of Technology, Mail Code 249-17, 1200 E. California Blvd, Pasadena, CA 91125, USA\\
$^{18}$European Southern Observatory, Alonso de Cordova 3107, Vitacura, Santiago, Chile\\
$^{19}$\\
$^{20}$Computational Engineering and Science Research Centre, University of Southern Queensland, Toowoomba, QLD 4350, Australia\\
$^{21}$Department of Astronomy, The Ohio State University, 140 West 18th Avenue, Columbus, OH 43210, USA\\
$^{22}$5 Inverness Way, Hillsborough, CA 94010, USA\\
$^{23}$AAVSO, 49 Bay State Rd., Cambridge, MA 02138, USA\\
$^{24}$Department of Astronomy \& Astrophysics, The Pennsylvania State University, 525 Davey Lab, University Park, PA 16802, USA\\
$^{25}$Center for Exoplanets and Habitable Worlds, The Pennsylvania State University, 525 Davey Lab, University Park, PA 16802, USA\\
$^{26}$Department of Physics and Astronomy, University of North Carolina at Chapel Hill, Chapel Hill, NC 27599-3255, USA\\
$^{27}$Department of Physics and Astronomy, Swarthmore College, Swarthmore, PA 19081, USA\\
$^{28}$Department of Physics, Westminster College, New Wilmington, PA 16172, USA\\
\end{footnotesize}
\vspace{0.3cm}\\
\bsp	
\label{lastpage}
\end{document}